\documentclass[twocolumn]{article}

\usepackage[utf8]{inputenc}
\usepackage{amsfonts,amssymb,amsmath}
\usepackage{graphicx}
\usepackage[authoryear]{natbib}
\usepackage{url}

\title{Analysis of $\sim10^6$ spiral galaxies from four telescopes shows large-scale patterns of asymmetry in galaxy spin directions}

\author{Lior Shamir \\ Kansas State University \\ Manhattan, KS 66506, USA \\ E-mail: lshamir@mtu.edu}

\date{}

\begin{document}

\maketitle

\begin{abstract}
The ability to collect unprecedented amounts of astronomical data has enabled the studying scientific questions that were impractical to study in the pre-information era. This study uses large datasets collected by four different robotic telescopes to profile the large-scale distribution of the spin directions of spiral galaxies. These datasets cover the Northern and Southern hemispheres, in addition to data acquired from space by the Hubble Space Telescope. The data were annotated automatically by a fully symmetric algorithm, as well as manually through a long labor-intensive process, leading to a dataset of nearly $10^6$ galaxies. The data shows possible patterns of asymmetric distribution of the spin directions, and the patterns agree between the different telescopes. The profiles also agree when using automatic or manual annotation of the galaxies, showing very similar large-scale patterns. Combining all data from all telescopes allows the most comprehensive analysis of its kind to date in terms of both the number of galaxies and the footprint size. The results show a statistically significant profile that is consistent across all telescopes. The instruments used in this study are DECam, HST, SDSS, and Pan-STARRS. The paper also discusses possible sources of bias, and analyzes the design of previous work that showed different results. Further research will be required to understand and validate these preliminary observations.
\end{abstract}


\section{Introduction}
\label{introduction}

While cosmological-scale isotropy is an elemental working assumption in cosmology, multiple observations using different probes have shown evidence of large-scale anisotropy. In addition to the anisotropy in the cosmic microwave background (CMB) \citep{eriksen2004asymmetries,cline2003does,gordon2004low,campanelli2007cosmic,zhe2015quadrupole,ashtekar2021cosmic,yeung2022directional}, large-scale anisotropy has been reported by analyzing the distribution of short gamma ray bursts \citep{meszaros2019oppositeness}, Ia supernova \citep{javanmardi2015probing,lin2016significance}, LX-T scaling \citep{migkas2020probing}, $H_o$ \citep{luongo2021larger}, dark energy \citep{adhav2011kantowski,adhav2011lrs,perivolaropoulos2014large,colin2019evidence}, high-energy cosmic rays \citep{aab2017observation}, quasars \citep{quasars,zhao2021tomographic}, and the frequency of galaxy morphology types \citep{javanmardi2017anisotropy}. Another specific observation that violates the cosmological isotropy assumption is the existence of the CMB Cold Spot \citep{cruz655non,mackenzie2017evidence,farhang2021cmb}. A correlation between higher $H_o$ and the CMB dipole has also been reported \citep{krishnan2021hints,luongo2021larger}.

The observations of large-scale anisotropy, and especially the anisotropy observed in the CMB, have led to models that shift from the standard cosmology. Explanations include primordial anisotropic vacuum pressure \citep{rodrigues2008anisotropic}, double inflation \citep{feng2003double}, moving dark energy \citep{jimenez2007cosmology}, contraction prior to inflation \citep{piao2004suppressing}, multiple vacua \citep{piao2005possible}, and spinor-driven inflation \citep{bohmer2008cmb}. Some explanations are related to the geometry of the Universe, such as ellipsoidal universe \citep{campanelli2006ellipsoidal,campanelli2007cosmic,campanelli2011cosmic,gruppuso2007complete,cea2014ellipsoidal}, and rotating universe \citep{godel1949example,ozsvath1962finite,ozsvath2001approaches,sivaram2012primordial,chechin2016rotation,camp2021}, where the large-scale anisotropy is expected to exhibit itself through a cosmological-scale axis. 

The existence of a cosmological-scale axis has also been linked to theories such as holographic big bang \citep{pourhasan2014out,altamirano2017cosmological}, and black hole cosmology \citep{pathria1972universe,easson2001universe,chakrabarty2020toy}, which is also related to flat space cosmology \citep{tatum2018flat,tatum2018clues}. These theories explain cosmic inflation without the need for dark energy. On the other hand, other cosmological models suggest that the possibility that dark energy itself is anisotropic cannot be ruled out \citep{adhav2011kantowski,adhav2011lrs}.


This study is focused on the probe of spin directions of spiral galaxies. A spiral galaxy is a unique extra-galactic object in the sense that its visual appearance is sensitive to the perspective of the observer. The spin directions of galaxies have been shown to be aligned within filaments on the cosmic web \citep{kraljic2021sdss}, but an alignment in the spin directions of galaxies was also observed when the galaxies are too far from each other to have gravitational interactions \citep{lee2019galaxy,lee2019mysterious}. A statistically significant correlation was also found between the spin direction of galaxies and cosmic initial conditions, proposing galaxy spin directions as a probe to study the early Universe \citep{motloch2021observed}. As these links are defined as ``mysterious'' \citep{lee2019mysterious}, the distribution of spin directions of spiral galaxies in the Universe is still unknown. 

In the past four decades, several studies provided evidence of non-random distribution in the spin directions of spiral galaxies  These research efforts started as early as the 1980's \citep{macgillivray1985anisotropy} with smaller datasets of several hundred spiral galaxies, and found non-random distribution with certainly of 92\% \citep{macgillivray1985anisotropy}. With the deployment of robotic telescopes that generate large astronomical databases, other studies using larger datasets of galaxies also showed evidence of non-random distribution \citep{longo2011detection,shamir2012handedness,shamir2013color,shamir2016asymmetry,shamir2017large,shamir2017photometric,shamir2017colour,shamir2019large,shamir2020pasa,shamir2020patterns,lee2019galaxy,lee2019mysterious,shamir2021particles,shamir2021large,shamir2022new}. On the other hand, other previous work argued that galaxy spin directions are distributed randomly \citep{iye1991catalog,land2008galaxy,hayes2017nature,iye2020spin}. These studies are described and analyzed in Section~\ref{previous_studies} of this paper. 

The disagreements between the results of different studies reinforce further analysis of the large-scale distribution of galaxy spin directions. Previous studies, whether argued that the distribution was random or not, used analyses such that all galaxies being analyzed were collected by the same instrument, which limited the size of these datasets. More importantly, analyzing data from a single instrument limits the size of the dataset footprint. To determine the nature of the distribution, it is therefore required to analyze large datasets of galaxies that cover a relatively large footprint of both the Northern and Southern hemispheres. The large number of galaxies can also enable sufficient statistical significance to determine whether the distribution of spin directions is random.
 
Here, data from several different instruments used in previous studies \citep{shamir2020pasa,shamir2020patterns,shamir2021large,shamir2022new} are combined into a single large dataset, providing a dataset of nearly 10$^6$ galaxies and a far larger footprint compared to any other dataset used for that purpose in the past. This ``meta analysis'' provides a more accurate profile compared to analyses based on datasets of smaller footprints. The profile observed with the combined dataset is also compared to the profiles observed with the datasets collected by single instruments. In addition to the analysis of a possible dipole axis done in \citep{shamir2022new}, the paper also analyzes quadrupole alignment.

\section{DATA}
\label{data}

The data were collected from four different telescopes. These include the Dark Energy Camera (DECam), Sloan Digital Sky Survey (SDSS), the Panoramic Survey Telescope and Rapid Response System (Pan-STARRS), and the Cosmic Assembly Near-infrared Deep Extragalactic Legacy Survey (CANDELS) sky survey imaged by Hubble Space Telescope (HST). The number of galaxies from each source is 807,898 from DECam, 33,028 from Pan-STARRS, 8,690 from HST, and 117,638 from SDSS. 

SDSS, Pan-STARRS, and the DESI Legacy Survey imaged by DECam are currently the largest and most productive digital sky surveys, with the largest footprints compared other Earth-based digital sky surveys. The data collected by these telescopes is publicly available, making these sky surveys suitable for this study. The sky surveys were also selected such that their combination cover both the Southern and Northern hemispheres. That provides a far larger footprint compared to any other previous study of this kind.  

In addition to the three Earth-based telescopes, another sky survey that was used was the Cosmic Assembly Near-infrared Deep Extragalactic Legacy Survey (CANDELS), imaged by HST. With the exception of the new James Webb Space Telescope (JWST), HST is the most productive space telescope working in the optical wavelength, providing substantial image data throughout its over two decades of service. As a space-based instrument, HST cannot match the vast bandwidth provided by the Earth-based sky surveys, and therefore the number of galaxies imaged by HST is substantially smaller compared to the other telescopes. The main advantage of HST is that it is not subjected to atmospheric effect, and therefore no unknown atmospheric effect can impact the analysis.

Because the footprints of the different sky surveys overlap, it is expected that some galaxies would be imaged by more than one telescope, and therefore can appear more than once after combining the four datasets into one. To avoid the same galaxy appearing in the dataset more than once, all objects in the combined dataset that had another object within less than 0.01$^o$ were removed. The exceptions are the galaxies imaged by HST, where the fields are dramatically smaller compared to the other sky surveys. HST galaxies are not bright enough to be imaged by the other sky surveys in a manner that allows to identify their spin direction, and are therefore not expected to be present in any of the other datasets. Combining all datasets provided a dataset of 958,841 different galaxies such that each galaxy appeared in the dataset exactly once. The specific datasets are described below.

\subsection{DECam data}
\label{desi}

The dark energy camera (DECam) of the Blanco 4 meter telescope is a powerful imaging instrument \citep{diehl2012dark,flaugher2015dark} capable of covering $\sim9\cdot10^3$ deg$^2$, mostly from the Southern hemisphere. The DECam data was retrieved through the DESI Legacy Survey \citep{dey2019overview}, which provides access to data acquired by multiple different instruments, including DECam.

The list of objects was retrieved from Data Release 8 of the DESI Legacy Survey, and included all objects imaged by DECam identified as galaxies, and had magnitude of less than 19.5 in either the g, r or z band. That provided a list of 22,987,246 objects identified as relatively bright galaxies. The images of these objects were downloaded by using the {\it cutout} service of the DESI Legacy Survey server. Each image is a 256$\times$256 JPEG image. The Petrosian radius was used to scale the image such that the object fits in the frame. To ensure full consistency of all images, all images were downloaded by the exact same computer. The process of downloading such a high number of galaxies lasted nearly nine months as described in \citep{shamir2021large}.

Due to the high number of galaxies, the annotation of the galaxies by their spin directions required an automatic process. Such process must be mathematically symmetric, and therefore needs to be based on clear define rules. While machine learning, and specifically convolutional neural networks, have been becoming very prevalent for solving problems in automatic analysis of galaxy images, they are based on complex data-driven automatically-generated rules that are very difficult to conceptualize. Because these rules are complex and non-intuitive, it is very difficult to verify that they are fully symmetric. For instance, neural networks are based on initial random weights that change during the training process, and differences between the images in the classes of the training set or even the order by which the images are being used in the training process can lead to differences in the neural network. That makes it virtually impossible to verify that the neural network is completely symmetric. Machine learning algorithms also tend to make forced choices. That is, even if the galaxy does not have a clear spin direction, the machine learning system will be forced to make a prediction. Slight asymmetries in the model that are very difficult to identify can therefore lead to small but consistent bias. More details about the possible consequences of using machine learning for this task are provided in Section~\ref{error}.

To have a fully symmetric annotation, the Ganalyzer algorithm was used \citep{shamir2011ganalyzer}. Ganalyzer is a model-driven algorithm that works according to mathematically defined rules, and it does not rely on data-driven rules or training data. Ganalyzer first transforms each galaxy image into its radial intensity plot. The radial intensity plot of an image is a 35$\times$360 image, such that the pixel $(x,y)$ in the radial intensity plot is the median value of the 5$\times$5 pixels around coordinates $(O_x+\sin(\theta) \cdot r,O_y-\cos(\theta)\cdot r)$ in the original galaxy image, where {\it r} is the radial distance measured in percentage of the galaxy radius, $\theta$ is the polar angle measured in degrees, and $(O_x,O_y)$ are the pixel coordinates of the center of the galaxy.

Because arm pixels are expected to be brighter than non-arm pixels at the same radial distance from the galaxy center, peaks in the radial intensity plot are expected to correspond to pixels on the arms of the galaxy at different radial distances from the center. Therefore, peak detection  \citep{morhavc2000identification} is applied to the lines in the radial intensity plot.

\begin{figure}[h]
\centering
\includegraphics[scale=0.23]{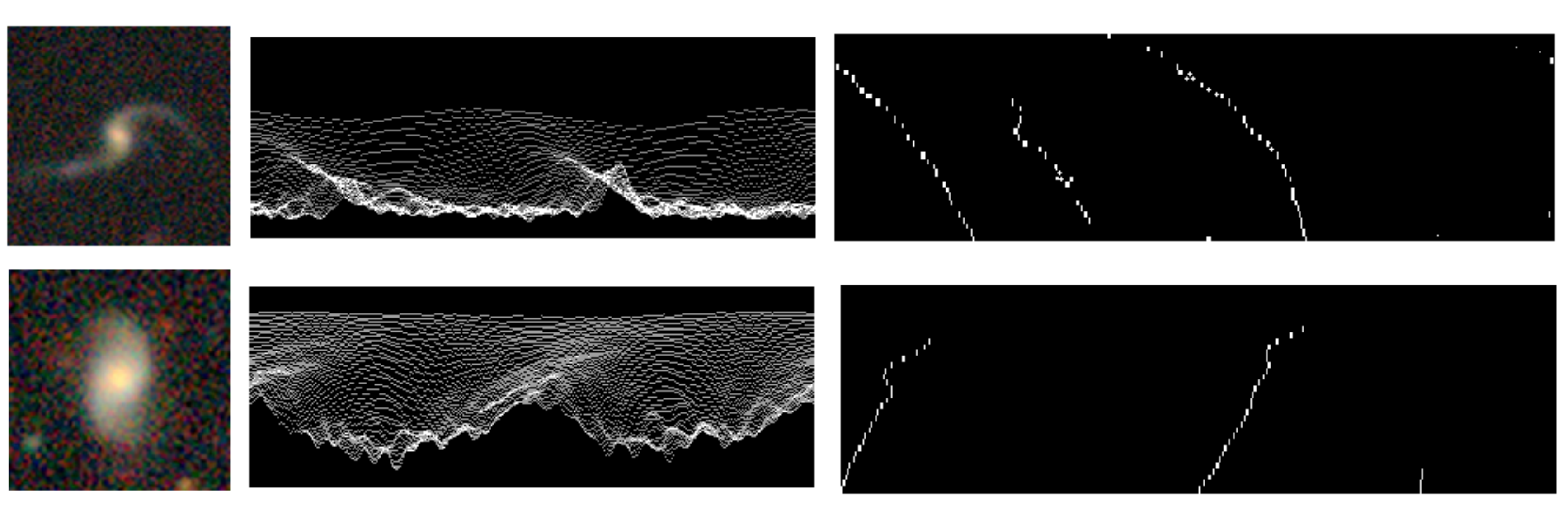}
\caption{Examples of the peaks of the radial intensity plots of different galaxy images. The direction of the lines generated by the peaks identifies the curves of the galaxy arms, and therefore can be used to determine the spin direction of the galaxy. The algorithm is fully symmetric, and is not based on complex non-intuitive data-driven rules commonly used in machine learning.}
\label{radial_intensity_plot}
\end{figure}

Figure~\ref{radial_intensity_plot} shows examples of two galaxies, their radial intensity plots, and the peaks identified in the radial intensity plots. As the figure shows, each arm is reflected by a vertical line of peaks. One of the galaxies has two arms, and therefore two vertical lines of peaks. The other galaxy has three arms, and therefore three lines of peaks. The direction towards which the peaks are aligned reflects the spin direction of the galaxy. More information about Ganalyzer can be found in \citep{shamir2011ganalyzer,dojcsak2014quantitative,shamir2017photometric,shamir2017colour,shamir2017large,shamir2019large,shamir2020patterns,shamir2021large,shamir2022new}.

\begin{figure}[h]
\centering
\includegraphics[scale=0.8]{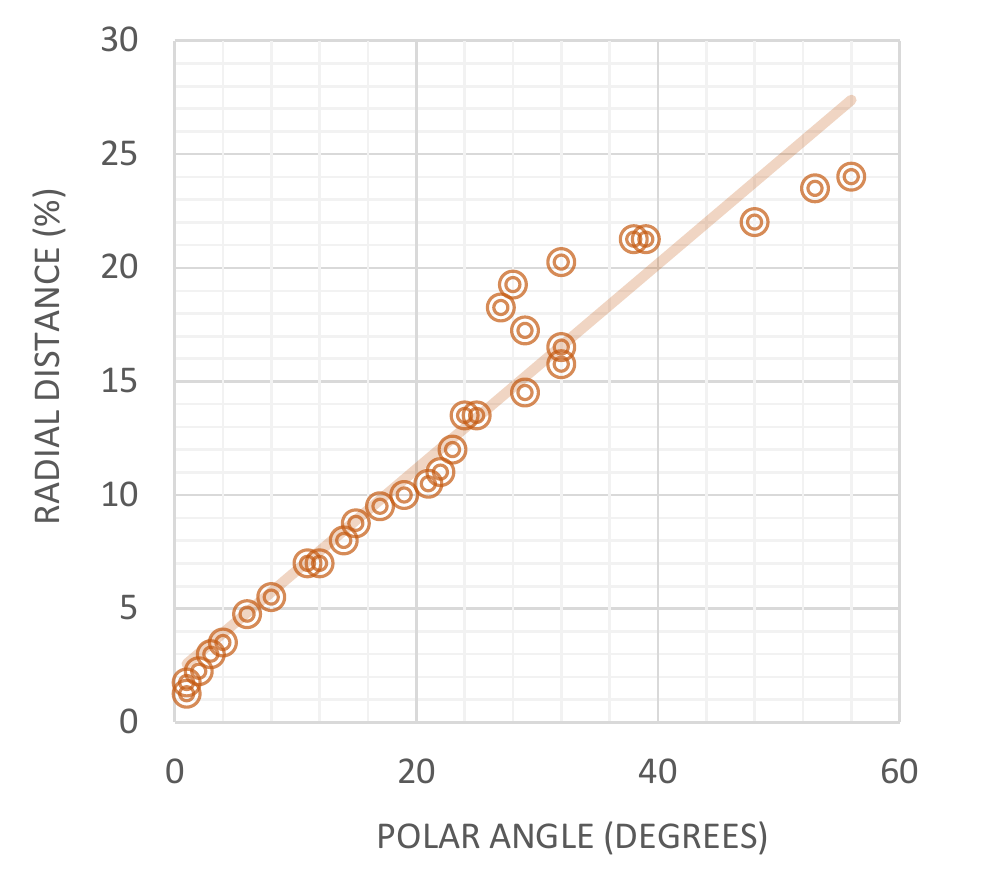}
\caption{Linear regression of the $(\theta,r)$ of the peaks of the left line (corresponding to the top arm of the galaxy) of the bottom galaxy in Figure~\ref{radial_intensity_plot}.}
\label{radial_intensity_plot_regression}
\end{figure}

The Cartesian coordinates of each peak $i$ are $(\theta_i,r_i)$, where $\theta_i$ is the polar angle of the peak $i$ compared to the galaxy center $(O_x,O_y)$, and $r_i$ is the radial distance from the galaxy center. The linear regression slope $\beta$ formed by these points is determined simply by the value of $\beta$ that satisfies $\min \Sigma_i (r_i- \beta \cdot \theta_i + \epsilon)^2$. If the slope $\beta$ is positive, the galaxy can be determined as spinning clockwise, while if $\beta$ is negative, the galaxy is a counterclockwise galaxy. For example, Figure~\ref{radial_intensity_plot_regression} shows the linear regression line of the peaks of the leftmost line of the bottom galaxy shown in Figure~\ref{radial_intensity_plot}. The slope of $\sim$0.45 is positive, and therefore the galaxy can be determined to be spinning clockwise. 

Not all $\sim2.2\cdot10^7$ galaxies are spiral galaxies, and not all spiral galaxies have identifiable spin direction. Therefore, the majority of the galaxies that were downloaded cannot be used for the analysis due to the fact that their spin direction cannot be identified. For that reason, galaxies that have at least 30 identified peaks in the radial intensity plot aligned at the same direction can be used. Galaxies that do not meet that threshold are rejected regardless of the sign of the linear regression of their peaks. That leaves a collection of 836,451 galaxies in the dataset that were assigned with an identifiable spin direction. Some of these galaxies are close satellite galaxies or other large extended objects inside a larger galaxy. Previous work suggested that the presence of duplicate objects can inflate the statistical significance \citep{iye2020spin}, and experiments by duplicating the objects artificially showed that an extremely high number of such objects can affect the statistical signal \citep{shamir2021particles}. A detailed discussion about the presence of duplicate objects is provided in Section~\ref{previous_studies}. To remove such objects, objects that have another object within less than 0.01$^o$ away were removed. That left 807,898 galaxies in the dataset. 

To test the consistency of the annotations, 200 random galaxies annotated as clockwise and 200 random galaxies annotated as counterclockwise were inspected manually, as was done in \citep{shamir2020patterns}. The visual inspection showed that none of the galaxies annotated by the algorithm as spinning clockwise was visually spinning counterclockwise, and none of the galaxies annotated as counterclockwise was by manual inspection spinning clockwise. Obviously, this small-scale test does not guarantee that no galaxies are missclassified, as the number of galaxies is too large to inspect manually. But the test suggests that the number of missclassified galaxies is expected to be small compared to the size of the data. More importantly, because the algorithm is symmetric, missclassified galaxies are expected to be distributed evenly between the different spin directions, and therefore cannot lead to asymmetry as explain theoretically and empirically in Section~\ref{error}.

To ensure the consistency of the galaxy annotation process, all images were analyzed on the exact same computer. That ensured that different system settings do not impact the analysis. Although there is no known computer system fault that can lead to differences in the annotation, full consistency was ensured by using just one computer system with a single processor. The annotation of the galaxies required 107 days of operation using a single Intel Xeon processor at 2.8 Ghz.

Tables~\ref{ra_distribution} and~\ref{dec_distribution} show the distribution of the galaxies by their right ascension and declination ranges, respectively. The DECam galaxies do not have redshift, and therefore the distribution of the redshift was determined by using a subset of 17,027 galaxies that had redshift in the 2dF data release \citep{cole20052df}. Table~\ref{z_distribution} shows the redshift distribution of the DECam galaxies.  


\begin{table}
\caption{The number of DECam galaxies in different 30$^o$ RA slices.}
\label{ra_distribution}
\centering
\begin{tabular}{lc}
\hline
RA           & \# galaxies  \\
(degrees) &                  \\
\hline
0-30     &  155,628  \\  
30-60  &  133,683 \\
60-90 &   80,134 \\ 
90-120 & 21,086 \\	
120-150 & 52,842 \\
150-180 &  59,660 \\
180-210 &  58,899 \\
210-240 & 58,112 \\
240-270 & 36,490 \\
270-300 & 2,602 \\
300-330 & 64,869 \\
330-360 & 83,893 \\
\hline
\end{tabular}
\end{table}

\begin{table}
\caption{The number of DECam galaxies in different declination ranges.}
\label{dec_distribution}
\centering
\begin{tabular}{lc}
\hline
Declination         & \# galaxies  \\
(degrees) &                  \\
\hline
-70 - -50    &  81,355  \\  
-50 - -30   &   123,972  \\
-30 - -10   &  121,656 \\ 
-10 - +10  &  236,740 \\	
+10 - + 30 & 203,562  \\
+30 - + 50 &  40,613 \\
\hline
\end{tabular}
\end{table}

\begin{table}
\caption{The number of DECam galaxies in different redshift ranges. The distribution is determined by a subset of 17,027 galaxies included in the 2dF data release.}
\label{z_distribution}
\centering
\begin{tabular}{lc}
\hline
z       & \# galaxies  \\
\hline
0-0.05    &  2,089  \\  
0.05-0.1   & 5,487    \\
0.1-0.15   &  4,226  \\ 
0.15 - 0.2  &  1,927 \\	
0.2-0.25 & 784 \\
0.25 - 0.3 &   621 \\
$>$0.3 &  1,893 \\
\hline
\end{tabular}
\end{table}

\subsection{SDSS data}
\label{sdss}

SDSS is an established digital sky survey that covers over 1.4$\cdot10^4$ deg$^2$, mostly in the Northern hemisphere. To study SDSS data, two datasets from SDSS that were used in previous studies were combined into one larger dataset. The two datasets were a dataset of $\sim6.4\cdot10^4$ galaxies with redshift \citep{shamir2019large,shamir2020patterns}, and another dataset of $\sim7.7\cdot10^4$ SDSS galaxies that do not have spectra. The preparation of these datasets is described in \citep{shamir2020patterns,shamir2021particles}. Both datasets were prepared by annotating the galaxies automatically as described in Section~\ref{desi}.

Since the two datasets are prepared from the same sky survey, their footprint naturally overlap, and some galaxies are included in both of datasets. To remove galaxies that appear in the combined dataset more than once, all objects in the combined dataset that had another object in the dataset within less than 0.01$^o$ were removed. That provided a combined dataset of 117,638 distinct galaxies. Table~\ref{ra_distribution_sdss} shows the RA distribution of the galaxies. More information about the distribution of the data and the way it was collected can be found in \citep{shamir2020patterns,shamir2021particles}.

\begin{table}
\caption{The number of SDSS galaxies in different RA 30$^o$ slices.}
\label{ra_distribution_sdss}
\centering
\begin{tabular}{lc}
\hline
RA           & \# galaxies  \\
(degrees) &                  \\
\hline
0-30     &  11,052  \\  
30-60  &  5,914 \\
60-90 &   1,520 \\ 
90-120 & 3,432 \\	
120-150 & 16,135 \\
150-180 &  19,083 \\
180-210 &  18,498 \\
210-240 & 18,443 \\
240-270 & 10,119 \\
270-300 & 631 \\
300-330 & 3,854 \\
330-360 & 8,957 \\
\hline
\hline
\end{tabular}
\end{table}


\subsection{Pan-STARRS data}
\label{pan-starrs}

The third digital sky survey used in this study is a dataset of $\sim3.3\cdot10^4$ galaxies from Pan-STARRS DR1 \citep{shamir2020patterns}. The initial set included 2,394,452 Pan-STARRS objects identified as extended sources by all color bands \citep{timmis2017catalog}. These galaxies were classified automatically by Ganalyzer \citep{shamir2011ganalyzer} as described in Section~\ref{desi}, and with more details in \citep{shamir2011ganalyzer,shamir2017photometric,shamir2017colour,shamir2017large,shamir2019large,shamir2020patterns}. That process provided 33,028 galaxies imaged by Pan-STARRS, and annotated by their spin direction. The distribution of the galaxies by their RA is shown in Table~\ref{PanSTARRS}. More information about the collection of the dataset and the distribution of the data can be found in \citep{shamir2020patterns}.

\begin{table}[ht]
{
\centering
\begin{tabular}{lc}
\hline
RA &   \# galaxies \\ 
\hline
0$^o$-30$^o$          &   3559  \\
30$^o$-60$^o$        &   2676  \\
60$^o$-90$^o$        &   1698 \\
90$^o$-120$^o$       &  1099 \\
120$^o$-150$^o$     &   3473 \\
150$^o$-180$^o$     &   5064 \\
180$^o$-210$^o$     &   5195 \\
210$^o$-240$^o$     &   4088 \\
240$^o$-270$^o$     &   1874 \\
270$^o$-300$^o$     &    429 \\
300$^o$-330$^o$     &   1074 \\
330$^o$-360$^o$     &   2799 \\
\hline
\end{tabular}
\caption{The number of Pan-STARRS galaxies in different 30$^o$ RA slices.}
\label{PanSTARRS}
}
\end{table}


\subsection{Hubble Space Telescope data}
\label{cosmos_data}

Although there is no atmospheric effect that can flip the spin pattern of a galaxy as observed from Earth, space-based observation can eliminate the possible impact of some unknown atmospheric effects that might make a galaxy spinning clockwise look as if it spins counterclockwise. For that purpose, a dataset of space-based observations was prepared from the Hubble Space Telescope (HST) Cosmic Assembly Near-infrared Deep Extragalactic Legacy Survey \citep{grogin2011candels,koekemoer2011candels}. The collection and preparation of that dataset is described in \citep{shamir2020pasa}.

The dataset was taken from several HST fields: the Cosmic Evolution Survey (COSMOS), the Great Observatories Origins Deep Survey North (GOODS-N), the Great Observatories Origins Deep Survey South (GOODS-S), the Ultra Deep Survey (UDS), and the Extended Groth Strip (EGS), providing an initial set of 114,529 galaxies \citep{shamir2020pasa}. The image of each galaxy was extracted by using {\it mSubimage} \citep{berriman2004montage}, and converted into 122$\times$122 TIF (Tagged Image File) image.

The initial number of galaxies imaged by HST is far smaller compared to other digital sky surveys such as the DECam survey. While the automatic analysis is fully symmetric, it also leads to the sacrifice of many galaxies that their spin direction cannot be identified with high certainty. Because the number of HST galaxies is smaller, the galaxies were annotated through a long labor-intensive process. During that process, a random half of the images were mirrored for the first cycle of annotation, and then all images were mirrored for a second cycle of annotation as described in \citep{shamir2020pasa} to offset the possible effect of perceptional bias. That provided a clean and complete dataset that is also not subjected to atmospheric effects \citep{shamir2020pasa}. The total number of annotated galaxies in the dataset was 8,690, and the distribution of the galaxies in the different fields is shown in Table~\ref{CANDELS}. Obviously, the HST galaxies are much more distant compared to the other telescopes, and has mean redshift of 0.58 \citep{shamir2020pasa}.

\begin{table}
\small
\caption{The number of galaxies in each of the five HST fields.}
\label{CANDELS}
\centering
\begin{tabular}{lccc}
\hline
Field & Field                        & \#           & Annotated \\ 
        & center                     &  galaxies   & galaxies    \\ 
\hline
COSMOS &  $150.12^o,2.2^o$ & 84,424 & 6,081 \\ 
GOODS-N & $189.23^o,62.24^o$ & 5,931 & 769 \\ 
GOODS-S & $53.12^o,-27.81^o$  & 5,024 & 540 \\ 
UDS        &  $214.82^o,52.82^o$   & 14,245 &  616 \\ 
EGS        &  $34.41^o,-5.2^o$    & 4,905    &  684 \\ 
\hline
\end{tabular}
\end{table}



The only parts of the sky that are covered by all four surveys are the HST fields. From these five fields, only the COSMOS field has a sufficient number of galaxies that allows certain statistical analysis. That field is also within the footprint of SDSS, Pan-STARRS, and DECam. Table~\ref{cosmos_sdss} shows the number of spiral galaxies spinning clockwise and the number of spiral galaxies spinning counterclockwise in the different datasets described in Section~\ref{data}.

The size of the COSMOS field is merely $\sim$2 square degrees. HST can naturally go deeper than any Earth-based sky survey, and therefore the number of galaxies in the COSMOS field is far larger than the number of galaxies in the same field in all the other sky survey. To have a sufficient number of galaxies that can allow statistical analysis, galaxies of the other digital sky survey were counted at the $10^o\times10^o$ field centered at the COSMOS field.

\begin{table}
\caption{The number of clockwise and counterclockwise galaxies in the COSMOS field, and in the $10^o\times10^o$ field of SDSS, Pan-STARRS, and DECam centered at COSMOS. The P values are the one-tail binomial probabilities \citep{wadsworth1960introduction} of having asymmetry equal or greater than the observed asymmetry when the probability of a galaxy to spin in a certain direction is 0.5.}
\label{cosmos_sdss}
\centering
\begin{tabular}{lllc}
\hline
Survey &        \# cw              & \# ccw & P \\
           &        galaxies          & galaxies                   &  \\
\hline
HST &  3,116 & 2,965 & 0.027 \\
Pan-STARRS     &  222  & 190 & 0.06 \\
SDSS  &  581  & 522 & 0.04 \\
DECam  & 2,640  & 2,498 & 0.025 \\
\hline
\end{tabular}
\end{table}

\section{Results}
\label{results}

\begin{figure}[h]
\centering
\includegraphics[scale=0.65]{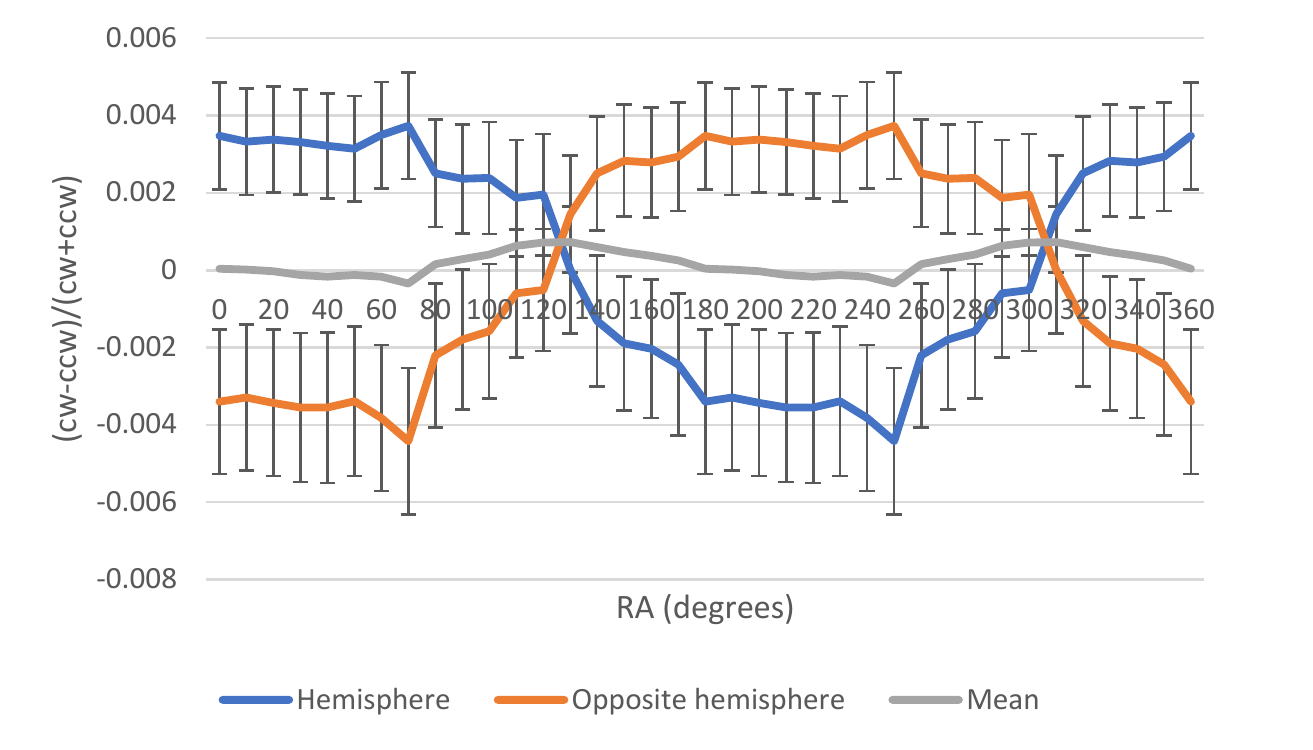}
\caption{The asymmetry between the number of galaxies that spin clockwise and the number of galaxies that spin counterclockwise in hemispheres centered at different RAs. The blue line shows the asymmetry in the hemisphere centered at the RA of the x-axis, and the orange line shows the same measurement in the opposite hemisphere. The error bars are the normal distribution standard error of $\frac{1}{\sqrt{n}}$, where {\it n} is the total number of galaxies in the hemisphere. The gray line shows the average of the asymmetry in both hemispheres.}
\label{hemispheres_all}
\end{figure}

The asymmetry {\it A} in each sky region is measured simply by $A=\frac{cw-ccw}{cw+ccw}$, where {\it cw} is the number of galaxies spinning clockwise, and {\it ccw} is the number of galaxies spinning counterclockwise. The error is determined by the normal distribution standard error of $\frac{1}{\sqrt{N}}$, where {\it N} is the total number of annotated galaxies in the sky region. Figure~\ref{hemispheres_all} shows the asymmetry between the number of clockwise galaxies and the number of counterclockwise galaxies in the 180$^o$ hemisphere centered at each RA, as well as the same measurement made in the opposite hemisphere. The figure shows that the asymmetry in one hemisphere is nearly exactly inverse to the asymmetry in the opposite hemisphere, and therefore the mean of the asymmetry observed in the opposite RA hemispheres is very close to zero. The figure also shows that the asymmetry is inverse in opposite hemispheres, and peaks around the hemisphere of $(130^o,310^o)$.

Table~\ref{hemispheres_180} shows a simple analysis by separating the galaxies by the RA range into the hemisphere $(130^o,310^o)$ and the opposite hemisphere $(<130^o \cup >310^o)$. The P value is the binomial probability to have such difference or stronger by chance when the probability for a galaxy to spin clockwise or counterclockwise is assumed at 0.5. Although the analysis is simple and does not account for differences in the declination, it still shows a higher number of galaxies spinning clockwise in one hemisphere, and a higher number of galaxies spinning counterclockwise in the opposite hemisphere.

\begin{table}
\small
\centering
\begin{tabular}{lcccc}
\hline
RA          & \# cw & \# ccw  & $\frac{cw-ccw}{cw+ccw}$  & P \\
              &   galaxies      &  galaxies                  &                                          & \\
\hline
$<130^o , > 310^o$        &   299,898     &  298,252 	  &   0.0028      &  0.017  \\   
$130^o-310^o$              &  179,765     & 180,926         &   -0.0032     &  0.026   \\
\hline
\end{tabular}
\caption{Number of clockwise and counterclockwise galaxies in the hemisphere $(130^o,310^o)$ and the opposite hemisphere $(<130^o \cup >310^o)$. The P value shows the binomial probability to have such distribution by chance if a galaxy has a probability of 0.5 to be assigned to a certain spin direction.}
\label{hemispheres_180}
\end{table}

\subsection{Analysis of a dipole axis in the distribution of galaxy spin directions}
\label{dipole}

The analysis shown above shows certain evidence that the sky can be separated into two hemispheres such that one has a higher number of clockwise galaxies and the opposite hemisphere has a higher number of counterclockwise galaxies. That analysis, however, is simplified by ignoring the declination of the galaxies, and the non-uniform distribution of the galaxy population imaged by the different sky surveys.

Following \cite{longo2011detection}, to test whether the distribution of the spin directions of the galaxies exhibits a dipole axis, $\chi^2$ statistics \citep{cochran1952chi2} was used to fit the galaxies in the datasets into the cosine of their angular distance from all possible integer $(\alpha,\delta)$ combinations. That was done by first assigning the galaxies with their spin direction $d$, which was {\it 1} if the spin direction of the galaxy is clockwise, and {\it -1} if the spin direction of the galaxy is counterclockwise. For each $(\alpha,\delta)$ combination, the angular distances $\phi$ between all galaxies in the dataset and $(\alpha,\delta)$ were computed.

Then, the cosines of the angular distances $\phi$ were $\chi^2$ fitted into $d\cdot|\cos(\phi)|$, where $d$ is the spin direction of the galaxy. The $\chi^2$ computed from each $(\alpha,\delta)$ integer combination was determined by Equation~\ref{chi2}
\begin{equation}
\chi^2_{(\alpha,\delta)}=\Sigma_i | \frac{(d_i \cdot | \cos(\phi_i)| - \cos(\phi_i))^2}{\cos(\phi_i)} | ,
\label{chi2}
\end{equation}
where $d_i$ is the spin direction of the galaxy (1 for clockwise and -1 for counterclockwise) {\it i}, and $\phi_i$ is the angular distance between galaxy {\it i} and $(\alpha,\delta)$.

To measure the statistical significance of the possible axis at $(\alpha,\delta)$, the $\chi^2_{(\alpha,\delta)}$ was also computed 1000 times such that in each run the galaxies were assigned with random spin directions. Using the $\chi^2_{(\alpha,\delta)}$ from 1000 runs, the mean $
\bar{\chi^2}^{random}_{(\alpha,\delta)}$ and standard deviation $\sigma^{random}_{(\alpha,\delta)}$ of the $\chi^2_{(\alpha,\delta)}$ when the spin directions are random was computed. Then, the statistical signal $\sigma_{(\alpha,\delta)}$ can be determined by Equation~\ref{sigma_difference}

\begin{equation}
\sigma_{(\alpha,\delta)}=\frac{|\chi^2_{\alpha,\delta}-\bar{\chi^2}^{random}_{(\alpha,\delta)}|}{\sigma^{random}_{(\alpha,\delta)}}
\label{sigma_difference}
\end{equation}

The $\sigma_{(\alpha,\delta)}$ difference between the $\chi^2$ computed with the real spin directions and the mean $\chi^2$ computed with the random spin directions was used to determine the $\sigma_{(\alpha,\delta)}$ of the $\chi^2$ fitness to occur by chance in each $(\alpha,\delta)$ combination. A detailed description of the analysis can be found in \citep{shamir2012handedness,shamir2019large,shamir2020pasa,shamir2020patterns,shamir2021particles}.

\begin{figure}[h]
\centering
\includegraphics[scale=0.14]{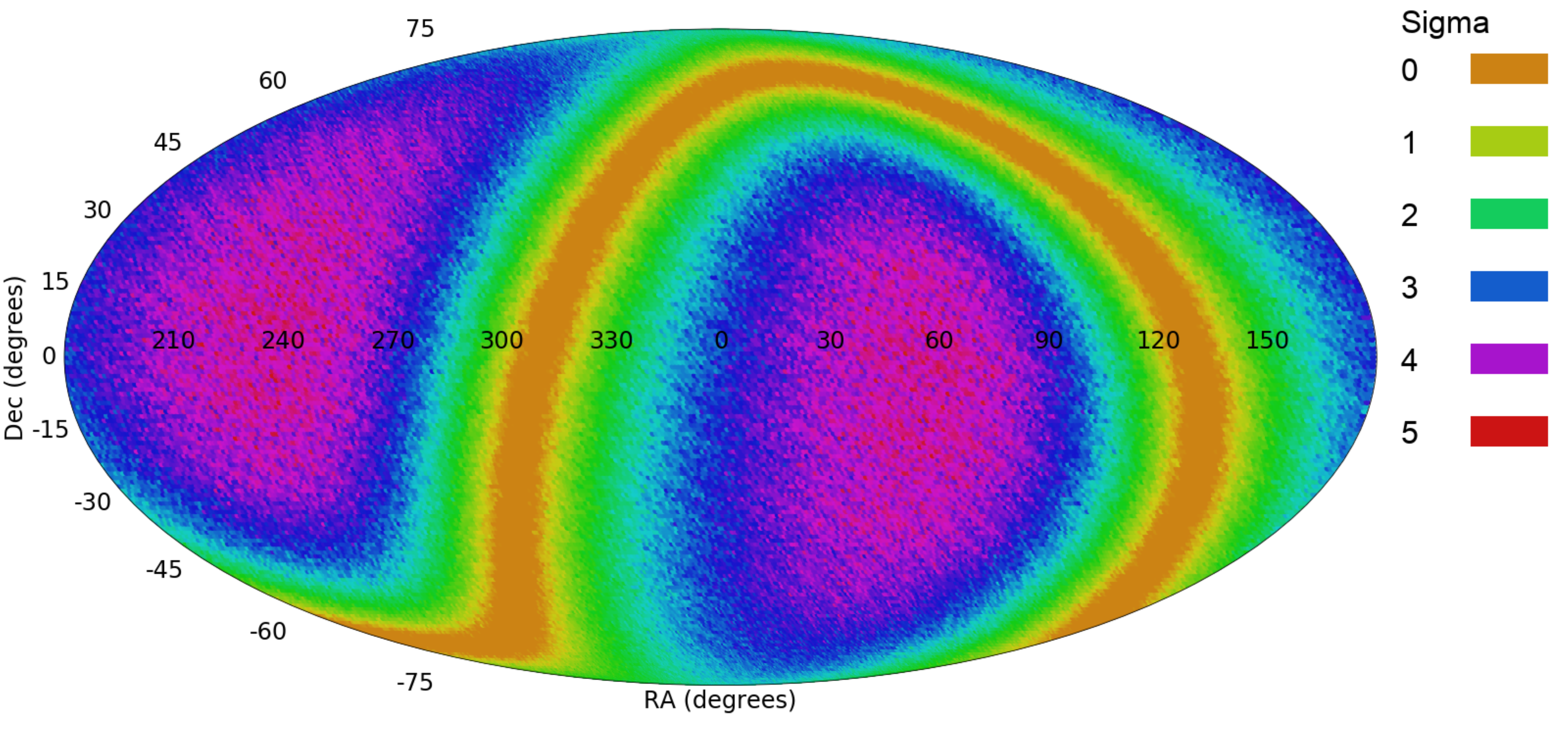}
\caption{The probability of a dipole axis in galaxy spin directions from all different $(\alpha,\delta)$ integer combination.}
\label{dipole_all}
\end{figure}

Figure~\ref{dipole_all} shows the probabilities $\sigma_{(\alpha,\delta)}$ of a dipole axis in different $(\alpha,\delta)$ coordinates, as defined by Equation~\ref{sigma_difference}. The figure shows a Mollweide projection of the $\sigma_{(\alpha,\delta)}$ computed in all integer $(\alpha,\delta)$ combinations by applying Equation~\ref{sigma_difference} to all possible $(\alpha,\delta)$. The most likely axis is identified at $(\alpha=47^o,\delta=-22^o)$, with probability of 3.7$\sigma$ to occur by chance. The 1$\sigma$ error of that axis is $(111^o > \cup > 344^o)$ for the RA, and $(-86^o,34^o)$ for the declination. Interestingly, the peak of the axis is nearly identical to the location of the CMB Cold Spot, at around $(\alpha=49,\delta=-19^o)$. 

While the proximity of the most likely axis to the CMB Cold Spot can definitely be considered a coincidence, the distribution of a large number of galaxies shows a statistically significant non-random distribution that forms a large-scale dipole axis. The observed presence of such Hubble-scale axis in the light of existing observations and current cosmological theories is discussed in Section~\ref{conclusion}.   

\begin{figure}[h]
\centering
\includegraphics[scale=0.14]{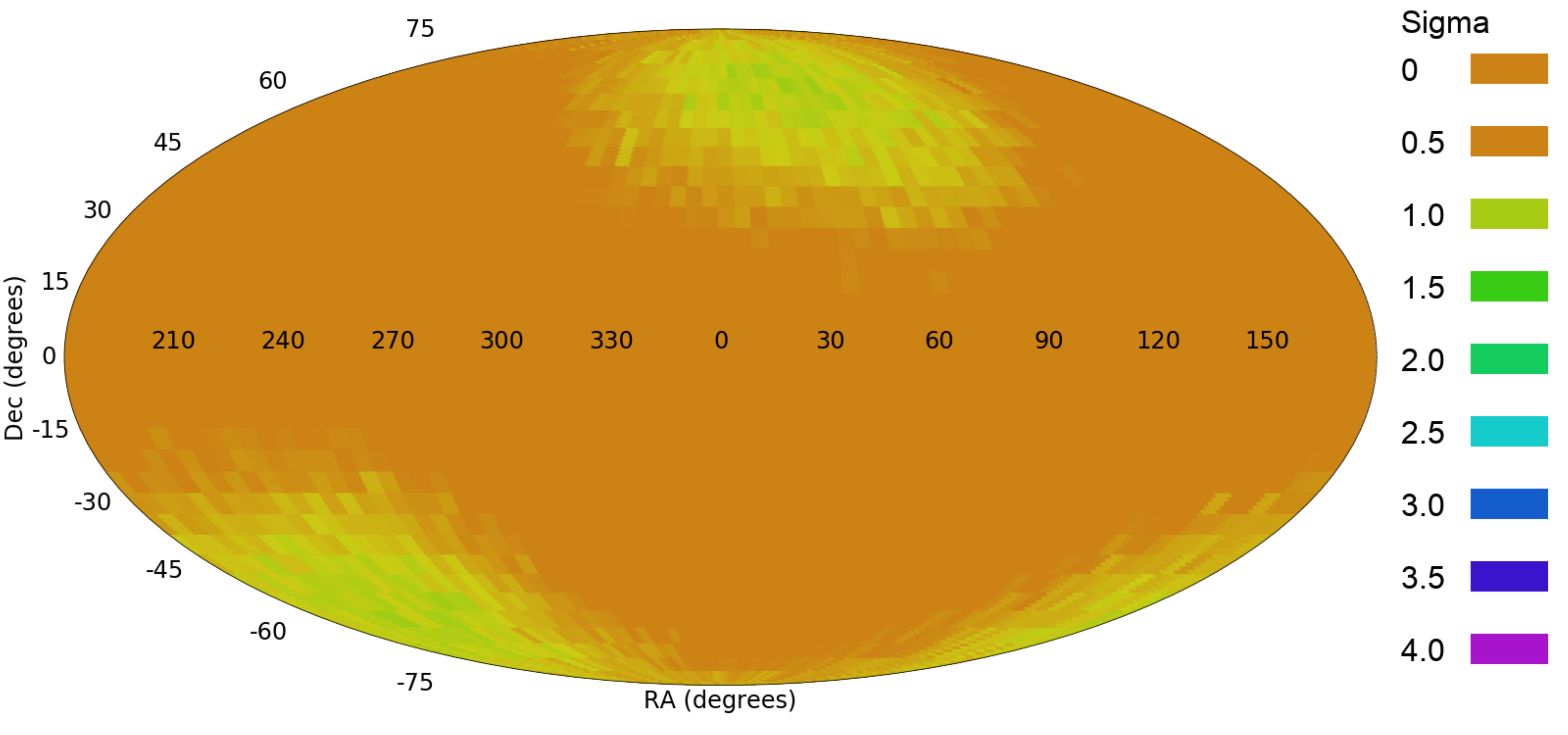}
\caption{The probability of a dipole axis from from different $(\alpha,\delta)$ when the galaxies are assigned with random spin directions.}
\label{dipole_all_random}
\end{figure}

Figure~\ref{dipole_all_random} shows the $\sigma_{(\alpha,\delta)}$  for all integer $(\alpha,\delta)$ combinations when the galaxies are assigned with random spin directions. That is the same analysis shown in Figure~\ref{dipole_all}, but the initial set of galaxies is assigned with random spin directions. As expected, the analysis showed no significant dipole axis when the galaxies are assigned with random spin directions. The strongest dipole axis had statistical significance of 0.81$\sigma$. That can be considered a control experiment, showing that the signal is present when the galaxies are assigned with their real spin directions, but becomes statistically insignificant when the spin directions are random.

\begin{figure*}[h!t]
\centering
\includegraphics[scale=0.14]{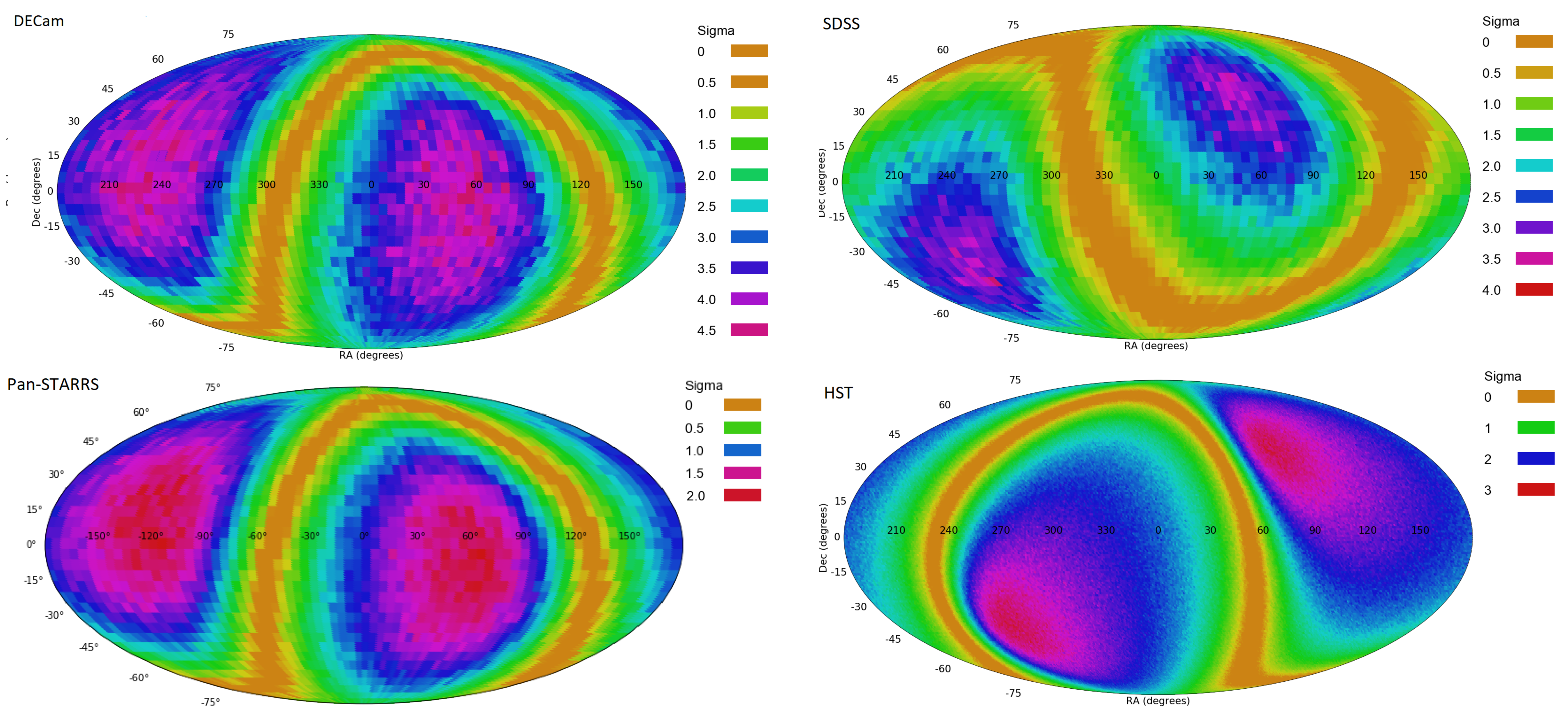}
\caption{The probability of a dipole axis in galaxy spin directions from different $(\alpha,\delta)$ combination in SDSS, Pan-STARRS, DECam, and HST.}
\label{dipole_datasets}
\end{figure*}

The data used to identify the most likely dipole axis was combined from several different sky surveys. By separating the data from each telescope it is possible to test whether the axis is consistent across different instruments \citep{shamir2022new}. Table~\ref{datasets} shows the results of applying the analysis to the data from each sky survey separately. As the table shows, the RA of the most likely dipole axes is aligned across all datasets, and well within the 1$\sigma$ error from each other. The differences in the declination are somewhat larger, but still within the 1$\sigma$ error. Because an Earth-based telescope can be either on the Northern or the Southern hemisphere, the declination range in the dataset of each telescope is not as broad as the RA range, and therefore the error in the declination is expected to be larger than the error in the RA. Figure~\ref{dipole_datasets} shows the probabilities of a dipole axis in the different $(\alpha,\delta)$ coordinates in the four sky surveys.

\begin{table}
\scriptsize
\centering
\begin{tabular}{|l|c|c|c|c|c|}
\hline
Dataset            &  RA             & Dec                    & $\sigma$    & RA error & Dec error     \\
                       &                    &                         &                   & (degrees)   &   (degrees)   \\
\hline
SDSS                       &  55$^o$      &   31$^o$   &    3.4     & (9,92)   &  (-25,77)            \\				
PanSTARRS            &  47$^o$       &   -11$^o$   &    1.87   &   (4,117)      &  (-73,40)              \\	
DECam  &   46$^o$      & -22$^o$   &    4.6   &  $(22,92)$  & $(-39,56)$ \\
HST                        &  78$^o$       &  47$^o$    &    2.8    &   (46,184)   & (-8,73)  \\
\hline
\end{tabular}
\caption{Most likely dipole axes from the data from the four digital sky surveys.}
\label{datasets}
\end{table}

\subsection{Analysis of a quadrupole axis in the distribution of galaxy spin directions}
\label{quadrupole}

Similarly to the analysis shown in Section~\ref{dipole}, the data was fitted into quadrupole alignment. That was done in the same manner of the dipole axis analysis, but by fitting $\cos(2\phi)$ to $d\cdot|\cos(2\phi)|$. Therefore, the analysis was the same as the analysis described in Section~\ref{dipole}, but when replacing Equation~\ref{chi2} with Equation~\ref{chi2_quad}

\begin{equation}
\chi^2_{(\alpha,\delta)}=\Sigma_i | \frac{(d_i \cdot | \cos(2\phi_i)| - \cos(2\phi_i))^2}{\cos(2\phi_i)} | ,
\label{chi2_quad}
\end{equation}

where $\phi$ is the angular distance between galaxy {\it i} and $(\alpha,\delta)$, and $d_i$ is the spin direction of galaxy $i$.

\begin{figure}[h]
\centering
\includegraphics[scale=0.15]{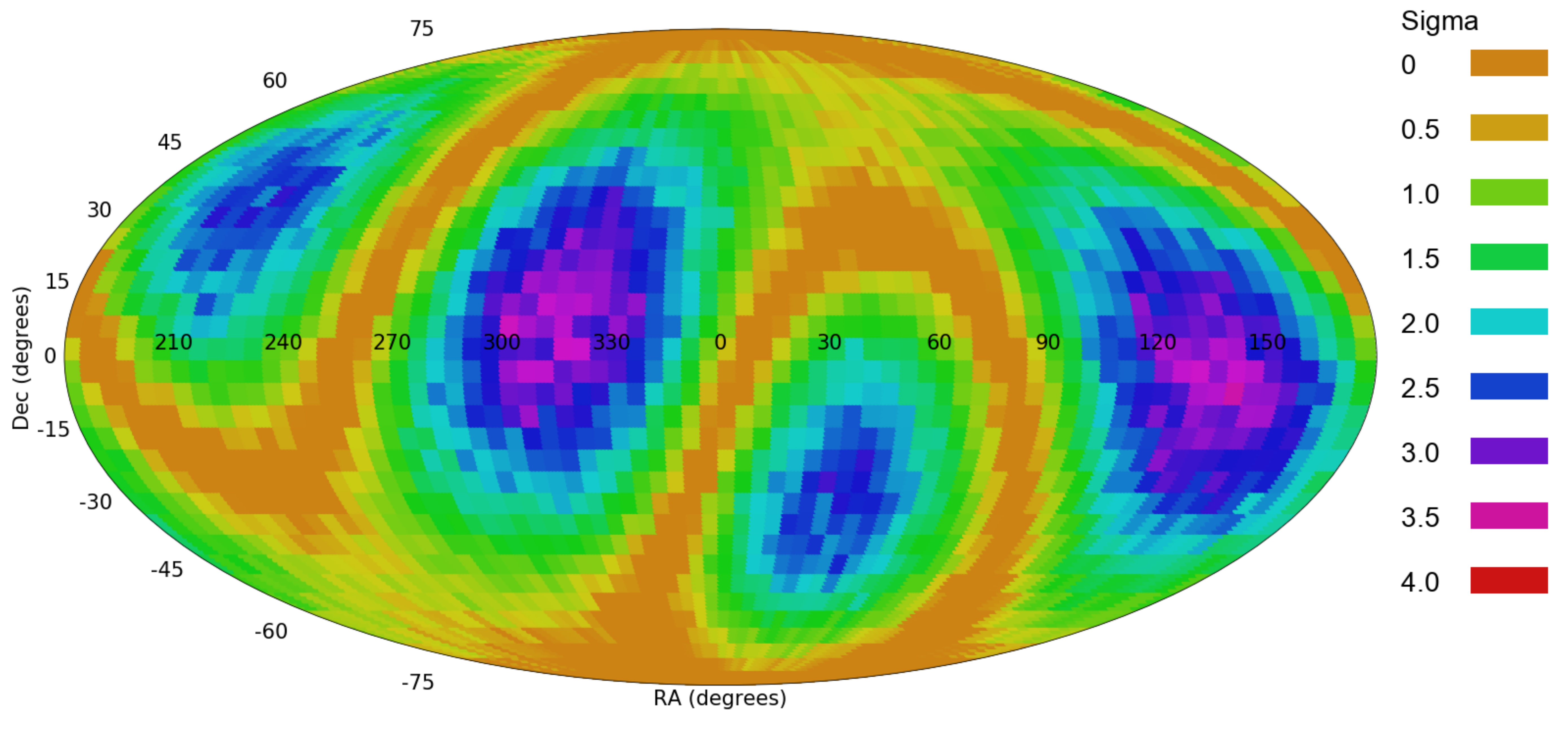}
\caption{The probability of a quadrupole axis in the galaxy spin directions at different $(\alpha,\delta)$.}
\label{quad_all}
\end{figure}

\begin{figure}[h]
\centering
\includegraphics[scale=0.15]{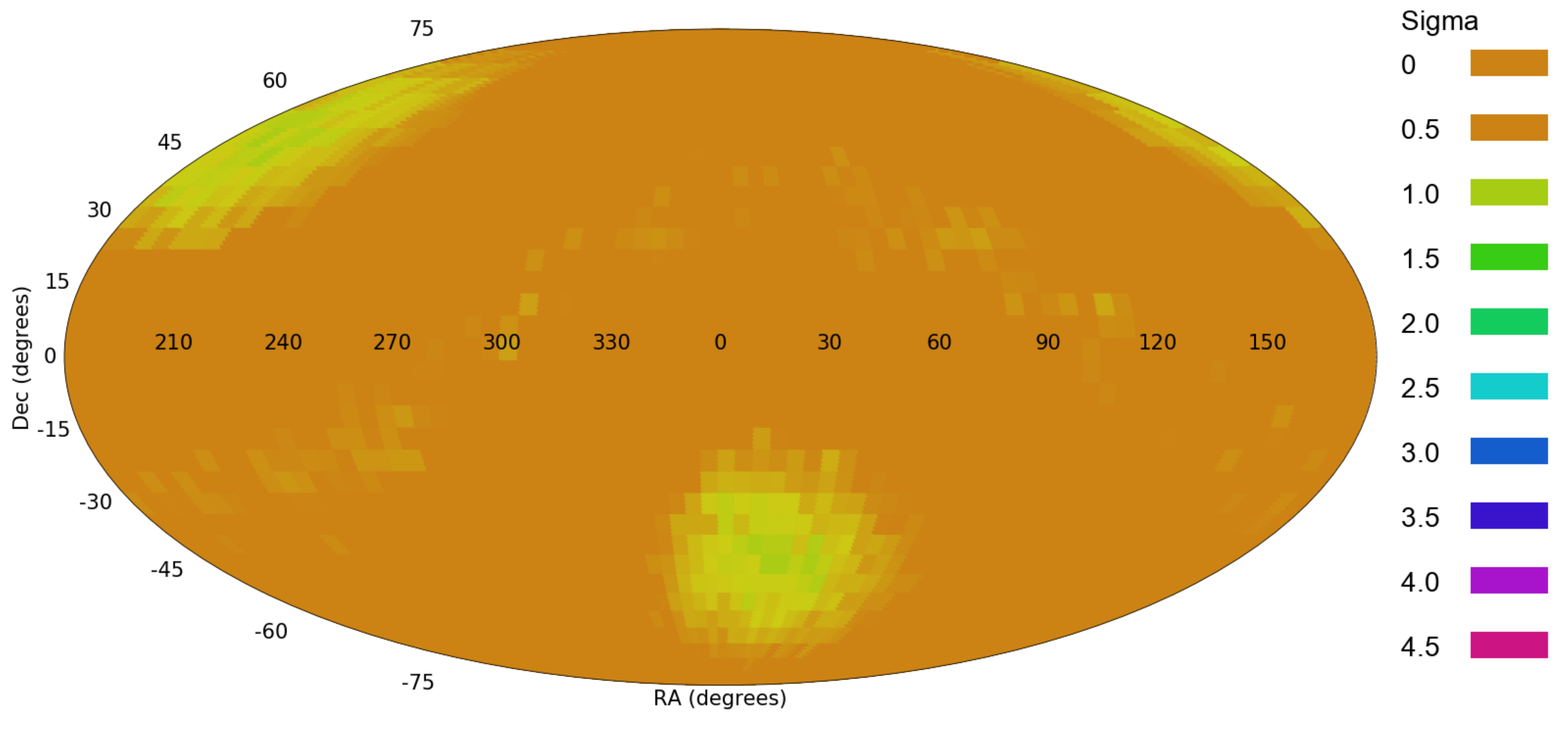}
\caption{The probability of a quadrupole axis in the galaxy spin directions when the galaxies are assigned with random spin directions.}
\label{quad_all_random}
\end{figure}

\begin{figure*}[h!t]
\centering
\includegraphics[scale=0.14]{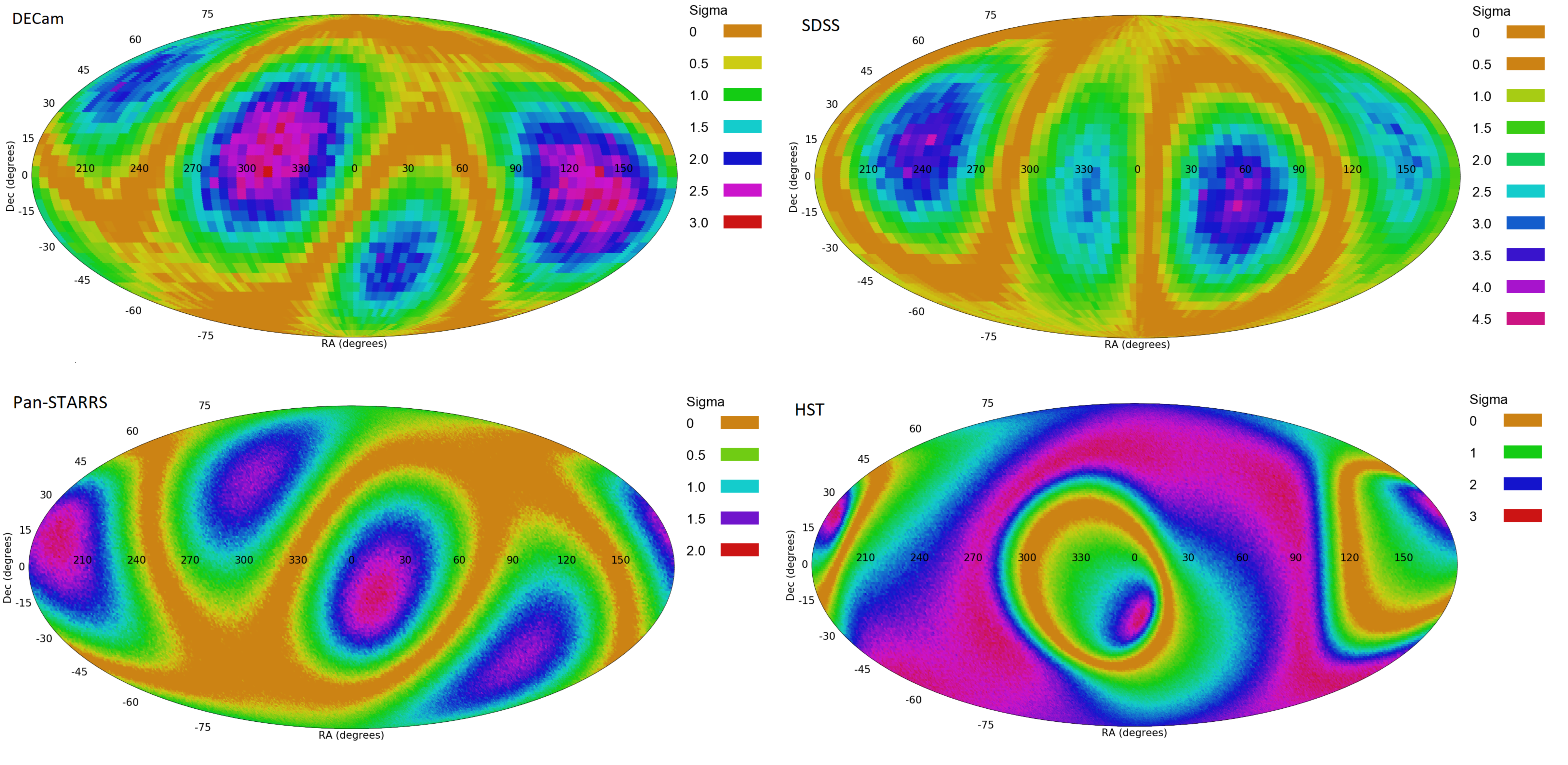}
\caption{The probability of a quadrupole axis in the data acquired by DECam, Pan-STARRS, SDSS, and HST. The HST dataset has a very small footprint, and therefore not expected to provide an informative quadrupole analysis.}
\label{quad_datasets}
\end{figure*}

Figure~\ref{quad_all} shows the probability of a quadrupole axis computed at different $(\alpha,\delta)$ coordinates. The analysis showed that the most probable quadrupole axes are at  $(\alpha=38^o,\delta=-28^o)$, with 2.9$\sigma$, and at $(\alpha=5^o,\delta=305^o)$, with probability of 3.2$\sigma$. Figure~\ref{quad_all_random} shows the same analysis such that galaxies are assigned with random spin directions, with a most probable axis of 0.76$\sigma$. Figure~\ref{quad_datasets} shows the same analysis with Pan-STARRS, SDSS, HST, and DECam data separately. HST has a very small footprint, and therefore could not be used effectively to analyze a quadrupole to show more than one peak.

\subsection{Analysis of galaxy spin directions around the location of the CMB Cold Spot}
\label{coldspot}

The analysis of a dipole axis done in Section~\ref{dipole} shows that the dipole axis peaks at close proximity to the location of the CMB Cold Spot, centered at around $(\alpha=48.77^o, \delta-19.58^o)$. While the alignment between the peak and the CMB Cold Spot can definitely be coincidental, the nature of the CMB Cold Spot is still poorly understood. Since both CMB and the spin directions of galaxies correlate with the initial conditions of the early Universe \citep{motloch2021observed}, a link between the CMB Cold Spot and galaxy spin should not be rules out.

To test the distribution of the galaxies around that part of the sky, the number of galaxies that spin clockwise was compared to the number of galaxies spinning counterclockwise in all telescopes. Since the CMB Cold Spot is relatively small, using just galaxies that appear in the field of the CMB Cold Spot will not provide a sufficient number of galaxies to make the comparison. Also, SDSS and Pan-STARRS do not have a very large galaxy population around that part of the sky. To use a larger field, the $40^o\times40^o$ sky region centered at the CMB Cold Spot was used. Table~\ref{coldspot_datasets} shows the number of clockwise and counterclockwise galaxies in each sky survey. Naturally, the HST dataset cannot be used for the analysis since there are no galaxies in that field that were imaged by that sky survey.

\begin{table}
\small
\caption{Number of clockwise and counterclockwise galaxies in the $40^o\times40^o$ region centered around the CMB Cold Spot. The P value is the binomial probability to have an equal or stronger distribution by mere chance.}
\label{coldspot_datasets}
\centering
\begin{tabular}{lcccc}
\hline
Dataset       & \# cw     & \# ccw    & $\frac{cw-ccw}{cw+ccw}$  & P \\
                  &  galaxies & galaxies  &   &  \\
\hline
All datasets &	54,850 & 53,723 & 0.0103	 & 0.0003	\\
SDSS          & 1,460 & 1,384 & 0.0267 & 0.074	\\
Pan-STARRS & 1,013 & 973 & 0.0201	 & 0.178	\\
DECam & 52,377 & 51,366 & 0.0097	 & 0.0008	\\
\hline
\end{tabular}
\end{table}

As the table shows, all sky surveys show a higher number of clockwise galaxies in that part of the sky. The asymmetry observed with SDSS and Pan-STARRS is not statistically significance, but also does not conflict with the asymmetry observed in the DECam data. SDSS shows difference that is marginally significant, with P$\simeq$0.07. But that can also be due to the fact that SDSS and Pan-STARRS have much less galaxies in that part of the sky. When combining SDSS and Pan-STARRS, the probability to have that asymmetry or stronger by chance is 0.046. While these results do not allow making a definite conclusion about a link between the CMB Cold Spot and galaxy spin directions, they provide certain a indication that can be explored by future empirical or theoretical studies.

\section{Possible errors}
\label{error}

One explanation to the observation would be an error in the analysis. This section discusses and explains several possible error, and shows that an error is unlikely.

\subsection{Error in the galaxy annotation algorithm}

An error in the annotation algorithm can obviously lead to asymmetry. However, multiple indications show that the asymmetry cannot be the result of an error in the classification algorithm. The algorithm is a model-driven symmetric algorithm with clear rules. It is not based on complex data-driven rules used by machine learning systems, which are virtually impossible to verify their symmetricity \citep{dhar2022systematic}. An experiment was performed by mirroring the galaxy images by using the {\it flip} command in the {\it ImageMagick} image analysis toolbox. As expected, mirroring the galaxies led to inverse asymmetry compared to the analysis with the original images.

Another evidence that the asymmetry is not driven by an error in the annotation algorithm is that the asymmetry changes between different parts of the sky, and inverse between opposite hemispheres. Since each galaxy is analyzed independently, a bias in the annotation algorithm is expected to be consistent throughout the sky, and it is not expected to flip in opposite hemispheres. The downloading of the images and the automatic analysis of the images were all done by the same computer, to avoid unknown differences between computers that can lead to bias or unknown differences in the way galaxy images are analyzed. 

Due to the theoretical and empirical evidence that the algorithm is symmetric, an error in the galaxy annotation is expected to impact clockwise and counterclockwise in a similar manner. If the galaxy annotation algorithm had a certain error in the annotation of the galaxies, the asymmetry {\it A} can be defined by Equation~\ref{asymmetry}.
\begin{equation}
A=\frac{(N_{cw}+E_{cw})-(N_{ccw}+E_{ccw})}{N_{cw}+E_{cw}+N_{ccw}+E_{ccw}},
\label{asymmetry}
\end{equation}
where $E_{cw}$ is the number of galaxies spinning clockwise incorrectly annotated as counterclockwise, and $E_{ccw}$ is the number of galaxies spinning counterclockwise incorrectly annotated as spinning clockwise. Because the algorithm is symmetric, the number of counterclockwise galaxies incorrectly annotated as clockwise is expected to be roughly the same as the number of clockwise galaxies missclassified as counterclockwise, and therefore $E_{cw} \simeq E_{ccw}$ \citep{shamir2021particles}. Therefore, the asymmetry {\it A} can be defined by Equation~\ref{asymmetry2}.

\begin{equation}
A=\frac{N_{cw}-N_{ccw}}{N_{cw}+E_{cw}+N_{ccw}+E_{ccw}}
\label{asymmetry2}
\end{equation}

Since $E_{cw}$ and $E_{ccw}$ cannot be negative, a higher rate of incorrectly annotated galaxies is expected to make {\it A} lower. Therefore, incorrect annotation of galaxies is not expected to lead to asymmetry, and can only make the asymmetry lower rather than higher.

\begin{figure}[h]
\centering
\includegraphics[scale=0.24]{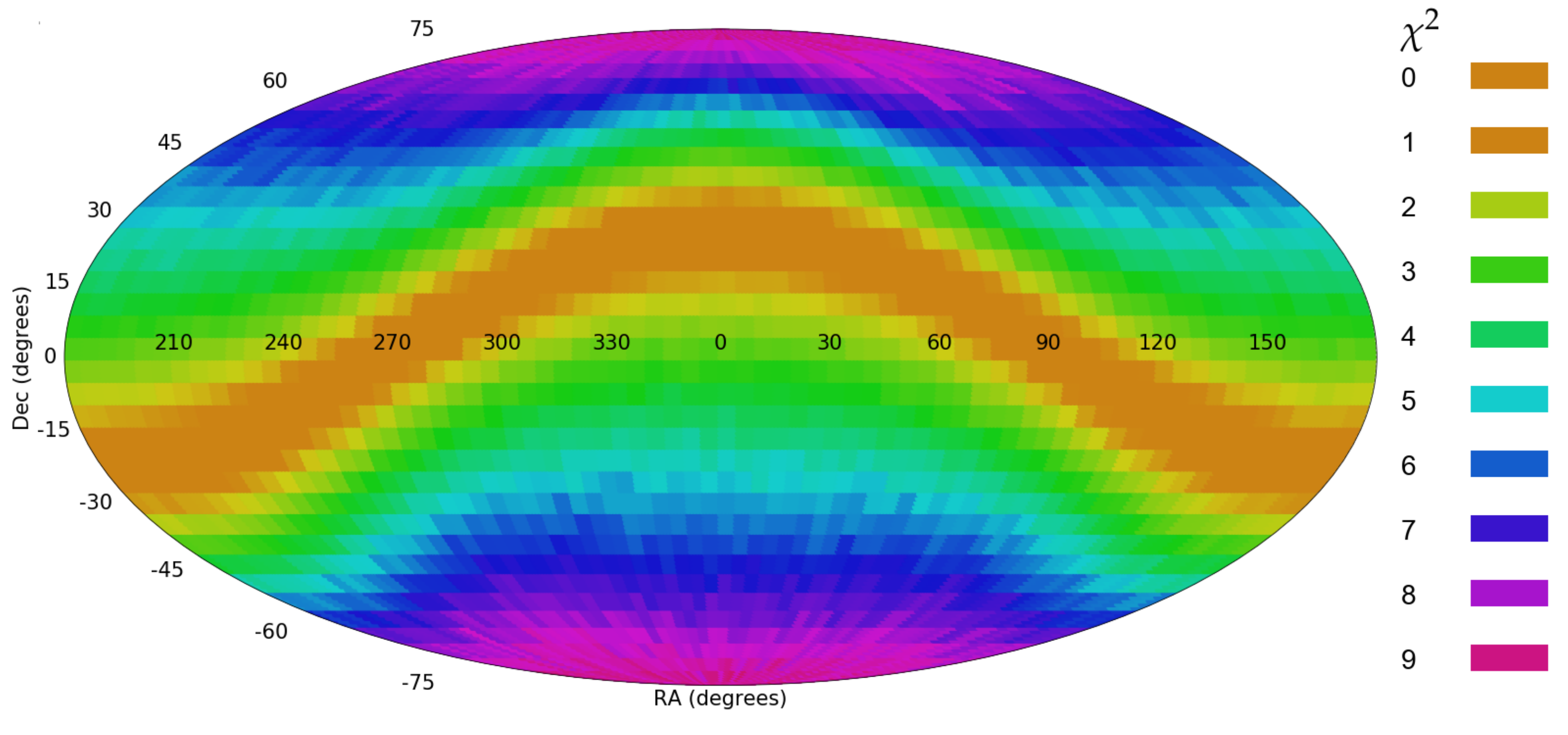}
\caption{Analysis of a dipole axis when using $7.7\cdot10^4$ SDSS galaxies, and assigning 2\% of the galaxies with clockwise spin direction regardless of their actual spin direction. The axis peaks exactly at the celestial pole, with a very strong statistical significance.}
\label{dipole_bias_2}
\end{figure}

An experiment \citep{shamir2021particles} of intentionally annotating some of the galaxies incorrectly showed that even when an error is added intentionally, the results do not change significantly even when as many as 25\% of the galaxies are assigned with incorrect spin directions, as long as the error is added to both clockwise and counterclockwise galaxies \citep{shamir2021particles}. But if the error is added in an asymmetric manner, even a small asymmetry of 2\% leads to a very strong asymmetry, and a dipole axis that peaks exactly at the celestial pole \citep{shamir2021particles}. Figure~\ref{dipole_bias_2} shows the results of analysis of $\sim7.7\cdot10^4$ SDSS galaxies after adding an artificial error of 2\%, meaning that a random 2\% of the galaxies are assigned with clockwise spin direction regardless of their real spin direction. The signal becomes immediately very strong, and peaks exactly at the celestial pole.

It should be mentioned that in one of the datasets used here, which is the dataset acquired by HST, the annotation was done manually, and without using any automatic classification. The galaxies imaged by HST were annotated manually, and the results are in agreement with the automatic annotation of galaxies imaged by SDSS, Pan-STARRS, and DECam.

\subsection{Bias in the sky survey hardware or photometric pipeline}

Autonomous digital sky surveys are some of the more complex research instruments, and involve sophisticated hardware and software to enable the collection, storage, analysis, and accessibility of the data. It is difficult to think of an error in the hardware or software that can lead to asymmetry between the number of clockwise and counterclockwise galaxies, but due to the complexity of these systems it is also difficult to prove that such error does not exist. That possible error is addressed here by using four different completely independent systems. DECam, SDSS, Pan-STARRS, and HST are completely independent from each other, and have different hardware and different photometric pipelines. As it is unlikely to have such bias in one instrument, it is very difficult to assume that all of these four instruments have such bias, and the profile of the bias is consistent across all of them.

\subsection{Cosmic variance}

The distribution of galaxies in the universe is not completely uniform. These subtle fluctuations in the density of galaxy population can lead to ``cosmic variance'' \citep{driver2010quantifying,moster2011cosmic}, which can impact measurements at a cosmological scale \citep{kamionkowski1997getting,amarena2018impact,keenan2020biases}. 

The probe of asymmetry between galaxies spinning in opposite directions is a relative measurement rather than an absolute measurement. That is, the asymmetry is determined by the difference between two measurements made in the same field, and therefore should not be affected by cosmic variance. Any cosmic variance or other effects that impacts the number of clockwise galaxies observed from Earth is expected to have a similar effect on the number of counterclockwise galaxies.

\subsection{Multiple photometric objects at the same galaxy}

In some cases, digital sky surveys can identify several photometric objects as independent galaxies, even in cases they are part of one larger galaxy. In the datasets used here all photometric objects that are part of the same galaxy were removed by removing all objects that had another object within 0.01$^o$. An exception is the HST galaxies, which are closer to each other due to the size of the field, but were inspected manually.

However, even if such objects existed in the dataset, they are expected to be evenly distributed between galaxies that spin clockwise and galaxies that spin counterclockwise, and therefore should not introduce an asymmetry. Experiments by using datasets of galaxies assigned with random spin directions and adding artificial objects to the galaxies showed that adding objects at exactly the same position of the original galaxies does not lead to signal of asymmetry \citep{shamir2021particles}. 

The experiments were made by using $\sim7.7\cdot10^4$ SDSS galaxies, and assigning the galaxies with random spin directions. Then, gradually adding more objects with the same location and spin directions as the galaxies in the original dataset, and the new artificial galaxies were assigned with the same spin direction as the galaxies in the original dataset \citep{shamir2021particles}. Adding such artificial galaxies did not lead to statistically significant signal.

\subsection{Atmospheric effect}

There is no known atmospheric effect that can make a galaxy that spin clockwise appear as if it spins counterclockwise. Also, because the asymmetry is always measured with galaxies imaged in the same field, any kind of atmospheric effect that affects galaxies the spin clockwise will also affect galaxies that spin counterclockwise. Therefore, it is unlikely that a certain atmospheric effect would impact the number of clockwise galaxies at a certain field, but would have different impact on galaxies spinning counterclockwise. In any case, one of the datasets used here is made of galaxies imaged by the space-based Hubble Space Telescope, and are therefore not subjected to any kind of atmospheric effect.

\subsection{Backward spiral galaxies}

In rare cases, the shape of the arms of a spiral galaxy is not an indication of the spin direction of the galaxy. An example is NGC 4622 \citep{freeman1991simulating}. A prevalent and systematically uneven distribution of backward spiral galaxies might indeed lead to asymmetry between the number of galaxies spinning clockwise and the number of galaxies spinning counterclockwise. For instance, if a relatively high percentage of galaxies that actually spin clockwise are backward spiral galaxies, it would have led to an excessive number of galaxies that seem to be spinning counterclockwise.

However, backward spiral galaxies are relatively rare. Also, these galaxies are expected to be distributed equally between galaxies that spin clockwise and galaxies that spin counterclockwise, and there is no indication of asymmetry between backwards spiral galaxies. Therefore, according to the known evidence, there is no reason to assume that the observations shown here are driven by backward spiral galaxies. The same can also apply to multi-spin galaxies \citep{rubin1994multi}, which are also rare, and should be equally distributed between both spin directions.

\section{Previous work showing different conclusions}
\label{previous_studies}

While several previous studies mentioned in Section~\ref{introduction} provided results suggesting that the large-scale distribution of galaxy spin directions is not necessarily random, other studies used similar approaches to reach opposite conclusions. It should be remembered that the null hypothesis is that the distribution of galaxy spin directions is random, and therefore could lead to the common bias known in science as ``confirmation bias'' \citep{hart2009}. This section analyzes these studies to identify reasons for the differences. 

An early attempt that showed random distribution was made by \cite{iye1991catalog}. In the absence of high-throughout digital sky surveys at the time, the analysis was based on a relatively small dataset of $\sim$6.5K galaxies. When assuming asymmetry of 1\% as shown here, 27,000 galaxies are needed to provide a one-tailed P value of 0.05. Even when assuming 2\% asymmetry, 7,000 galaxies are needed to provide one-tailed binomial distribution probability of P$\simeq$0.048. Therefore, a dataset of $\sim$6K galaxies is too small to provide a statistically significant observation of the asymmetry.

Another study that used manual annotation of galaxies was based on crowdsourcing done by unprofessional volunteers through Galaxy Zoo \citep{land2008galaxy}. The approach had the advantage of using a large number of volunteers to increase the bandwidth of the annotation. Its main downside was that the annotations were subjected to human bias \citep{land2008galaxy}. That led to inaccuracy of the annotations, but more importantly, the bias of the annotations was systematic. Because the attempt to use crowdsourcing for that task was first of its kind, the presence and dominance of the perceptual bias was not known when the experiment was designed, and therefore the galaxy images were not mirrored randomly to offset for the bias. 

After applying a process of data correction by mirroring the images of a small subset of the galaxies, the results using the mirrored and original galaxy images showed an asymmetry of 1\%-2\%, That can be seen in Table 2 in \citep{land2008galaxy} that summarizes the results of the small subset of galaxies that were corrected for the human bias by mirroring the galaxies. The table shows that when mirroring the galaxies the number of galaxies annotated as counterclockwise was reduced by $\sim$1.5\% (from 6.032\% counterclockwise galaxies to 5.942\% mirrored clockwise galaxies), while the number of galaxies annotated as clockwise increased by $\sim$2\% (from 5.525\% clockwise galaxies to 5.646\% mirrored counterclockwise galaxies). That asymmetry is similar in direction and magnitude to the asymmetry shown in \citep{shamir2020patterns}. The observation reported in \citep{shamir2020patterns} is the most suitable comparison since it also analyzes SDSS galaxies with spectra, and therefore the footprint and distribution of the galaxies is similar.

Due to the corrections of the human annotators, the number of galaxies used in \citep{land2008galaxy} for the analysis became much smaller than the initial number of galaxies, and the asymmetry was determined to be statistically insignificant. However, the results also do not disagree with the results shown with SDSS galaxies here and in \citep{shamir2020patterns}. The magnitude and direction of the asymmetry observed with Galaxy Zoo data are aligned with the results observed with SDSS data used here, although there is no statistical significance to neither accept nor reject that agreement. It has also been proposed that non-random distribution of the spin directions of the galaxies annotated by Galaxy Zoo cannot be ruled out \citep{motloch2021observational}.

A study that used automatic annotation of the spin directions of spiral galaxies was by \cite{hayes2017nature}. The abstract suggests that ``when viewed across the entire GZ1 sample (and by implication, the Sloan catalogue), the winding direction of arms in spiral galaxies as viewed from Earth is consistent with the flip of a fair coin``. That conclusion certainly conflicts with the results shown here. To understand the reason for that conflict, one might need to pay close attention to the details of the experimental design. The explanation to the absence of asymmetry can be explained by one sentence in Section 4.1, that explains the implementation of the annotation algorithm used to determine the spin direction of the galaxies: ``We choose our attributes to include some photometric attributes that were disjoint with those that Shamir (2016) found to be correlated with chirality, in addition to several SPARCFIRE outputs with all chirality information removed.``. 

That is, to create a machine learning algorithm that can determine the spin direction of galaxies, \cite{hayes2017nature} removed attributes that correlate with the spin direction asymmetry that were reported in \citep{shamir2016asymmetry}. Naturally, when removing specifically the attributes that correlate with the asymmetry in spin direction, the machine learning algorithm produced a dataset that is fully symmetric, and aligned with random distribution of the spin directions. 

The attributes that correlate with galaxy spin direction asymmetry identified in \citep{shamir2016asymmetry} do not have an obvious direct link to galaxy spin direction. \cite{hayes2017nature} do not provide a scientific motivation for removing these attributes, and it seems that the decision to remove them was observational rather than scientific, with the goal of removing ``bias''. Ignoring specifically the attributes that correlate with galaxy spin direction asymmetry naturally removed the asymmetry and led to a system that provided a dataset that showed no asymmetry. That experiment, however, could be biased by the selection of the attributes.

When using all attributes, the asymmetry between the number of clockwise and counterclockwise galaxies was with statistical significance of 2.52$\sigma$, as specified in Table 2 in \citep{hayes2017nature}. That distribution is not necessarily random, and in fact agrees with the results shown here more than it agrees with the null hypothesis. These results are also in agreement with previous analysis of SDSS galaxies as reported in \citep{shamir2020patterns}. 

As explained in \citep{hayes2017nature}, the random distribution was observed only after removing the specific attributes that are known to correlate with the asymmetry between clockwise and counterclockwise galaxies. Ignoring these attributes naturally led to a random distribution, but since certain specific attributes were intentionally ignored, that distribution may or may not reflect the distribution of the galaxy spin directions in the real sky. In any case, the experimental design according which all attributes that are known to reflect asymmetry between clockwise and counterclockwise galaxies are ignored naturally leads to an algorithm that produces a randomly distributed dataset. When not removing these attributes, the observed distribution was 2.52$\sigma$, which is not necessarily random.

Another study that showed opposite results used the dataset of \citep{shamir2017photometric} and suggested that the asymmetry is the result of `'duplicate objects” in the dataset \citep{iye2020spin}. When removing the ``duplicate objects'' to create a `'clean” dataset, the signal drops to 0.29$\sigma$. As the abstract claims ``The actual dipole asymmetry observed for the “cleaned” catalog is quite modest, $\sigma_D$ = 0.29''.

However, the dataset used in \citep{shamir2017photometric} was used for photometric analysis. No claim for the presence or absence of any kind of dipole axis was made in \citep{shamir2017photometric}, and no such claim about that dataset was made in any other paper. When using that dataset for analyzing the distribution of the galaxy population, photometric objects that are part of the same galaxy become ``duplicate objects”. But as mentioned above, there was no attempt to study the presence or absence of a dipole axis with that dataset \citep{shamir2017photometric}, and no claim about any kind of dipole axis formed by that dataset was made in \citep{shamir2017photometric} or in any other paper.

But the more interesting question is why a ``clean” dataset showed random distribution of the galaxy spin directions. The answer can be found in a sentence in Section 3 of the paper: ``The second sample we studied is a volume-limited sample retaining 111,867 spirals with measured redshift (Paul et al. 2018) in the range 0.01 $<$z $<$0.1.''

As also explained in \citep{shamir2021large}, the signal of 0.29$\sigma$ reported in the abstract of the paper was observed with the ``second sample'', where the redshift of the galaxies is limited to z$<$0.1. As shown in \citep{shamir2019large,shamir2020patterns}, when limiting the redshift to lower redshift ranges of z$<$0.15, the distribution of galaxy spin directions is random. That is also shown in Tables 3, 5, 6, and 7 in \citep{shamir2020patterns}. These tables show random distribution in lower redshift ranges. Therefore, the random distribution in $z<0.1$ reported in \citep{iye2020spin} is completely expected, and in full agreement with previous work \citep{shamir2019large,shamir2020patterns}. The paper \citep{iye2020spin} does not provide a scientific motivation for limiting the redshift to 0.1. 

More importantly, the ``measured redshift” used to determine the dipole axis is in fact the photometric redshift from the catalog of \citep{paul2018catalog}. In the analysis shown in this paper and in all previous work, the position of each galaxy is determined by its RA and declination, which are considered accurate measurements. In \citep{iye2020spin}, however, the analysis is three dimensional, and the position of each galaxy is determined by its RA, declination, and distance. The distance $d^i$ of each galaxy $i$ is computed by $d^i=cz^i/H_{o}$ , where {\it c} is the speed of light, $H_{o}$ is the Hubble Constant, and $z^i$ is the redshift of galaxy {\it i}. Because the vast majority of the galaxies in that dataset do not have spectra, the distance was determined by using the photometric redshift. The photometric redshift is a highly inaccurate, ambiguous (in the sense that one galaxy can have multiple different photometric redshifts), and systematically biased. The error of the photometric redshift used for the analysis is $\sim$18.5\% \citep{paul2018catalog}, which is far greater than the $\sim$1-2\% signal of asymmetry reported here and in previous work. The very substantial error of the photometric redshift is therefore expected to weaken the signal. Because the photometric redshift is determined by complex pattern recognition rules, the systematic bias of the photometric redshift might also impact the results in a manner that is difficult to predict. For these reasons, the photometric redshift is not a sound probe for analyzing subtle anisotropies in the large-scale structure. The 3D analysis such that the distance was determined by using the photometric redshift is therefore expected to lead to random distribution. Indeed, all results shown in \citep{iye2020spin} are completely different from the results shown here or in all previous work. For instance, the ``unclean" dataset showed a dipole axis with statistical strength of 4.00$\sigma$ when the dataset was limtied to $z_{phot}<0.1$ \cite{iye2020spin}, while the only previous attempt to limit to lower redshifts showed no statistically significant dipole axis in that redshift range \citep{shamir2020patterns}.

When using the photometric redshift for determining the position of the galaxies, the observed signal when not limiting the redshift range is 1.29$\sigma$ \citep{iye2020spin}. Since the photometric redshift is highly inaccurate, and in fact its inaccuracy is far greater than the expected signal, it is likely that the photometric redshift leads to a substantially weaker statistical signal. Indeed, an analysis by the National Astronomical Observatory of Japan showed that when using basic statistics where the photometric redshift is not used, the distribution of the galaxy spin directions in that dataset is not random \citep{Fukumoto2021}. The analysis is as follows:

Table~\ref{hemispheres} shows the number of clockwise and counterclockwise SDSS galaxies in the hemisphere centered at RA=160$^o$, and in the opposite hemisphere in the exact same dataset used in \citep{iye2020spin}. Statistically significant signal is observed in the hemisphere centered at 160$^o$. The asymmetry in the opposite and less populated hemisphere is not statistically significant. But because it has more counterclockwise galaxies than clockwise galaxies, it is also not in conflict with the distribution in the hemisphere centered at (RA=160$^o$) for forming a dipole axis. Because there are two hemispheres, the two-tailed probability needs to be corrected to $\sim$0.01. That simple analysis provides certain evidence that the distribution in the specific dataset of SDSS galaxies used in \citep{iye2020spin} might not be random.

\begin{table}
\scriptsize

\begin{tabular}{|l|c|c|c|c|}
\hline
Hemisphere & \# cw          & \# ccw         &  $\frac{\#Z}{\#S}$  & P  \\ 
   (RA)         &                        &                         &                     & (one-tailed)  \\ 
\hline
$70^o-250^o$               & 23,037 & 22,442   &   1.0265   &  0.0026 \\ 
$>250^o \cup <70^o$   &  13,660 &  13,749  &   0.9935   &  0.29   \\ 
\hline
\end{tabular}
\caption{The number of clockwise and counterclockwise spiral galaxies in the exact same dataset used in \citep{iye2020spin} in the hemisphere centered at $\alpha=160^o$, and in the opposite hemisphere, centered at $\alpha=340^o$. The P values are based on binomial distribution such that the probability of a galaxy to spin clockwise or counterclockwise is 0.5.}
\label{hemispheres}
\end{table}

A Monte Carlo simulation was applied such that each galaxy was assigned with a random spin direction, and a search was applied to test whether any two hemispheres have an asymmetry described in Table~\ref{hemispheres} or stronger. Out of 10,000 runs, 155 runs provided a distribution that could be divided into two hemispheres with equal or stronger asymmetry to the two hemispheres shown in Table~\ref{hemispheres}. That also shows results that might not necessarily be random.

\begin{figure}[h]
\centering
\includegraphics[scale=0.14]{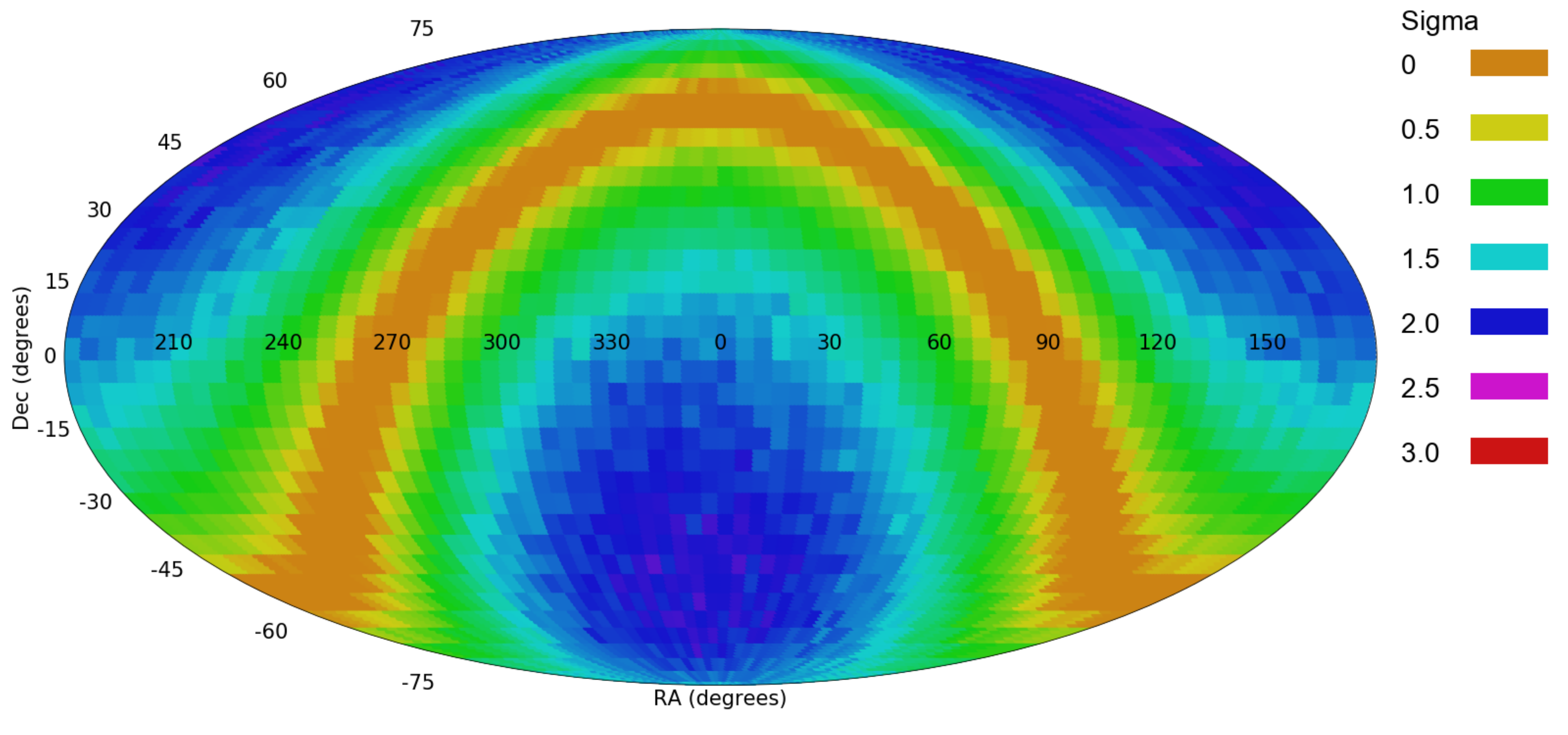}
\caption{The $\chi^2$ probability of a dipole axis in the spin directions of the galaxies from different $(\alpha,\delta)$ combinations in the exact same dataset used by \citep{iye2020spin}.}
\label{dipole1}
\end{figure}

Figure~\ref{dipole1} shows the statistical significance of the dipole axis from each possible pair of integer ($\alpha$,$\delta$) when using the exact same dataset used by \citep{iye2020spin}, but without using the photometric redshift to determine the position of each galaxy. The dataset contains 72,888 galaxies, and available at \url{https://people.cs.ksu.edu/~lshamir/data/assym_72k/}. The most likely location of the dipole axis is identified at $(\alpha=165^o,\delta=40^o)$, and the statistical signal of the axis is 2.16$\sigma$. That statistical signal is not necessarily random, and does not conflict the observations shown in Section~\ref{results}. Since the dataset of \citep{shamir2017photometric} contains bright galaxies (i magnitude$<$18, Petrosian radius $<$5.5'), it is expected that these galaxies are also of lower redshift, and therefore the asymmetry is expected to be weaker as shown in \citep{shamir2020patterns}. In any case, the statistical strength of the asymmetry is $>2\sigma$, and cannot be considered necessarily random.

\begin{figure}[h]
\centering
\includegraphics[scale=0.14]{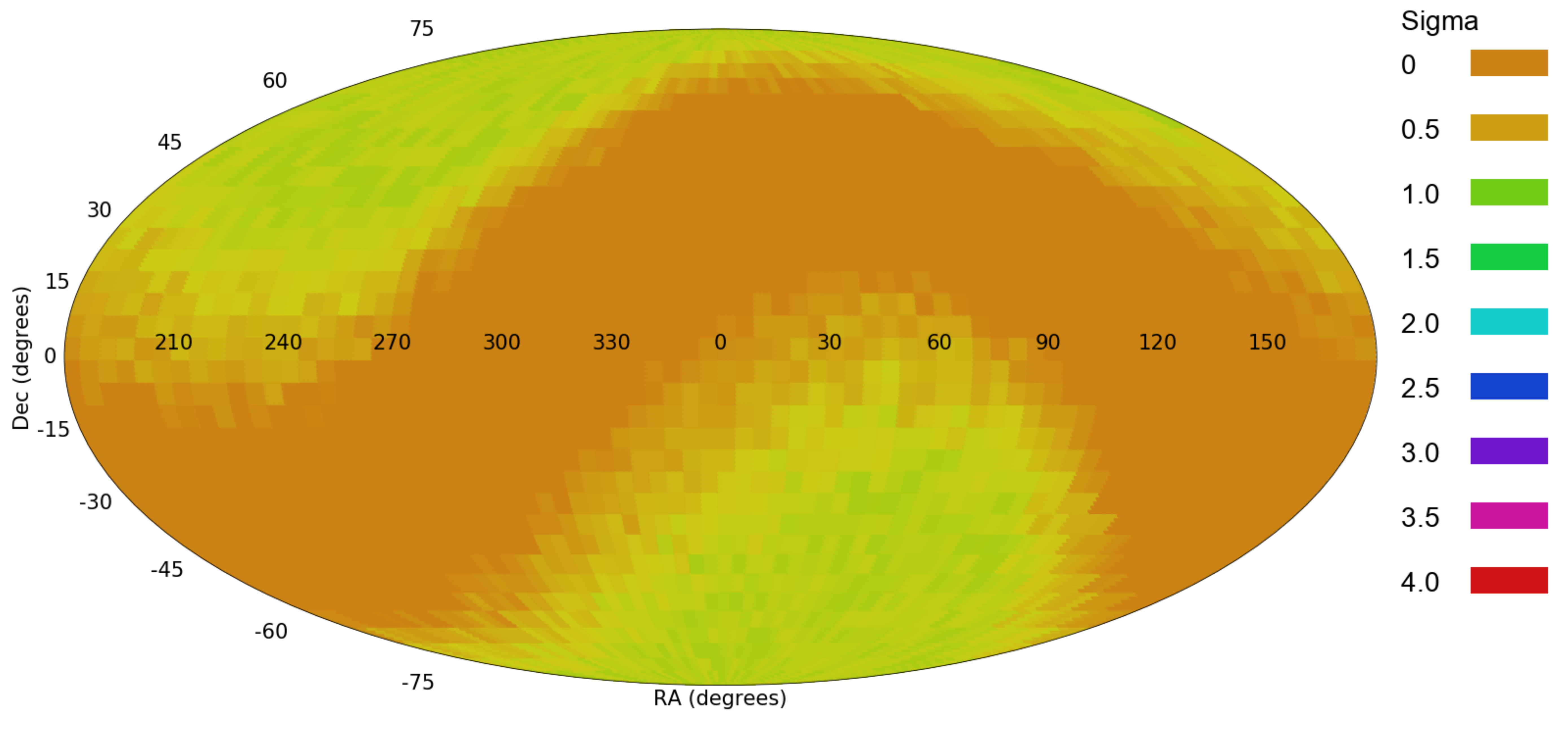}
\caption{The $\chi^2$ probability of the dipole axis when the galaxies are assigned with random spin directions.}
\label{dipole_random}
\end{figure}

Figure~\ref{dipole_random} shows the likelihood of the dipole axis when the galaxies are assigned with random spin direction, showing much lower probability of $<1\sigma$.

\begin{figure}[h]
\includegraphics[scale=0.14]{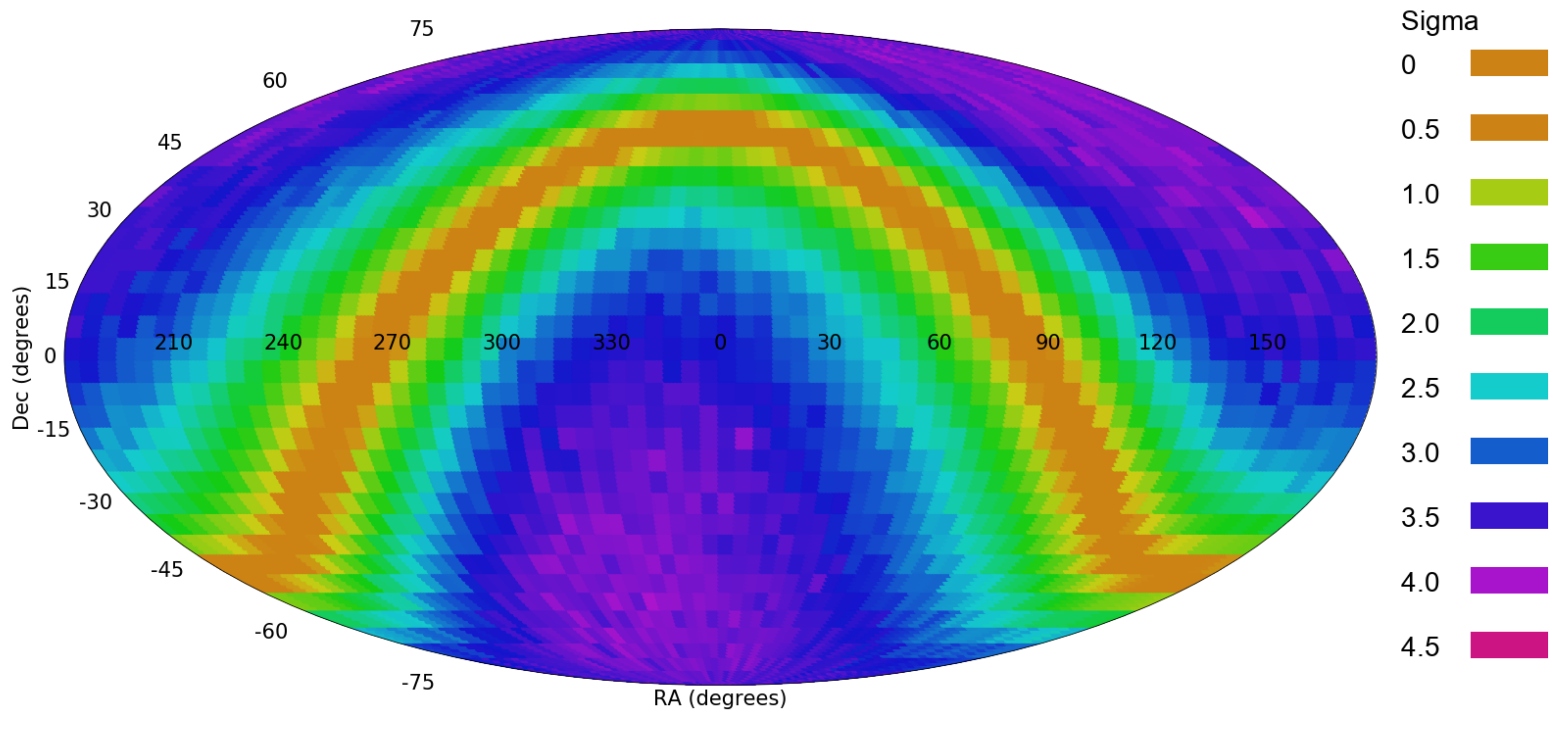}
\caption{The $\chi^2$ probability of the dipole axis when using the  dataset of 13,440 manually annotated SDSS galaxies.}
\label{sdss2_dipole}
\end{figure}

These results also can be also compared to the results when using the dataset that \citep{hayes2017nature} used to determine and remove attributes that correlate with the galaxy spin direction asymmetry. The dataset contains 13,440 galaxies that were annotated manually, and available at \url{https://people.cs.ksu.edu/~lshamir/data/assym}. Figure~\ref{sdss2_dipole} shows the probabilities of a dipole axis to peak at the different $(\alpha,\delta)$ combinations. The figure shows a similar profile to the profile shown in Figure~\ref{dipole1}, and non-random distribution.

\section{Discussion}
\label{conclusion}

Autonomous digital sky surveys powered by robotic telescopes have allowed the collection of unprecedented amounts of astronomical data, enabling to address research questions that were not addressable in the pre-information era. The question addressed here is the large-scale distribution of the spin directions of spiral galaxies as observed from Earth. Multiple previous experiments have shown that the distribution of spin directions of spiral galaxies as observed from Earth might not be random, and might form patterns at scales far larger than any known cluster or supercluster \citep{longo2011detection,shamir2012handedness,shamir2013color,shamir2016asymmetry,shamir2017large,shamir2017photometric,shamir2017colour,shamir2019large,shamir2020pasa,shamir2020patterns,lee2019galaxy,lee2019mysterious,shamir2021particles,shamir2021large,shamir2022new}.  Analysis of galaxies with spectra and separating the galaxies to different redshift ranges showed that asymmetry is weak at low redshifts, but increases gradually as the redshift gets higher \citep{shamir2020patterns,shamir2022new}.

This study shows the most comprehensive and largest analysis of its kind to date. The analysis uses several different telescopes systems with different photometric pipelines. The analysis covers the Northern hemisphere, the Southern hemisphere, and uses both space-based and ground-based instruments. Each dataset is analyzed independently, and without using any assumptions from other datasets. While the telescopes cover different parts of the sky, based on completely different hardware, use different photometric pipelines, and the data were annotated using different methods, all telescopes show very similar patterns of the asymmetry. The agreement in the results from the different telescope systems and the different parts of the sky provides an indication of consistency, showing that the observations do not necessarily depend on a specific dataset. As discussed in Section~\ref{introduction}, other datasets collected in the past four decades also showed asymmetry, although the small size of these datasets did not allow to profile the distribution within statistical significance. Section~\ref{previous_studies} analyzed studies that suggested random distribution, and showed that these results do not necessarily conflict with non-random distribution.

Some first attempts to study the distribution of the spin directions of galaxies were based on manual annotation of the galaxies \citep{land2008galaxy,longo2011detection}, showing evidence of non-random distribution. 
By using a fully symmetric algorithm, much larger databases can be analyzed without the possible effect of human perception \citep{shamir2012handedness,shamir2013color,shamir2016asymmetry,shamir2019large,shamir2020pasa,shamir2020patterns}. The application of the automatic annotation to data acquired by several different telescopes showed similar profiles of distribution. The distribution can be fitted to dipole or quadrupole alignment with probability far higher than mere chance.



Studies with smaller datasets of galaxies showed non-random spin directions of galaxies in filaments of the cosmic web \citep{tempel2013evidence,tempel2013galaxy,kraljic2021sdss}. Other studies showed alignment in the spin directions even when the galaxies are too far from each other to interact gravitationally \citep{lee2019galaxy,lee2019mysterious}, unless assuming modified Newtonian dynamics (MOND) gravity models that explain longer gravitational span \citep{sanders2003missing,darabi2014liquid,amendola2020measuring}. It should be mentioned that the physics of galaxy rotation is still not fully understood, and it is still not clear why and how galaxies spin. While the common theory that can explain the anomaly in the galaxy rotation curve \citep{rubin1983rotation} is the existence of dark matter, there is still no certain proof that dark matter indeed exists \citep{giagu2019wimp}. More recent observations showed that cosmic filaments also spin, and the origin of their spin can be explained by angular momenta originating from the Universe initial conditions \citep{sheng2022spin}.  

Other observations of large-scale alignment in spin directions were observed with quasars \citep{hutsemekers2014alignment}. Position angle of radio galaxies also showed large-scale consistency of angular momentum \citep{taylor2016alignments}. These observations agree with observations made with datasets such as the Faint Images of the Radio Sky at Twenty-centimetres (FIRST) and the TIFR GMRT Sky Survey (TGSS), showing large-scale alignment of radio galaxies \citep{contigiani2017radio,panwar2020alignment}. Large-scale clustering suggesting evidence for axis alignment were also observed in {\it Fermi} blazars \citep{marcha2021large}.

In addition to the empirical observations, simulations of dark matter also showed links between spin directions and the large-scale structure \citep{zhang2009spin,libeskind2013velocity,libeskind2014universal}. The magnitude of the correlation has been associated with the color and stellar mass and the galaxies \citep{wang2018spin}, and that association was linked to halo formation \citep{wang2017general}, leading to the contention that the spin direction in the halo progenitors is related to the large-scale structure of the early universe \citep{wang2018build}. 

The large-scale analysis of spin directions done here shows evidence of dipole and quadrupole large-scale alignment. The results with DECam data agree with previous results using \citep{shamir2020pasa,shamir2020patterns,shamir2021particles}. The observation of a large-scale axis has been proposed in the past by analyzing the cosmic microwave background (CMB), with consistent data from the Cosmic Background Explorer (COBE), Wilkinson Microwave Anisotropy Probe (WMAP) and Planck \citep{abramo2006anomalies,mariano2013cmb,land2005examination,ade2014planck,santos2015influence,gruppuso2018evens,yeung2022directional}. Observations also showed that the axis formed by the CMB temperature is aligned with other cosmic asymmetry axes such as dark energy and dark flow \citep{mariano2013cmb}. Other notable statistical anomalies in the CMB are the quandrupole-octopole alignment \citep{schwarz2004low,ralston2004virgo,copi2007uncorrelated,copi2010large,copi2015large}, the asymmetry between hemispheres \citep{eriksen2004asymmetries,land2005examination,akrami2014power}, point-parity asymmetry \citep{kim2010anomalous,kim2010anomalous2}, and the CMB Cold Spot. If these anomalies are not statistical fluctuations \citep{bennett2011seven}, they can be viewed as observations that disagree with $\Lambda$CDM \citep{bull2016beyond}.

The most likely dipole axis identified using the spin directions of galaxies shown here peaks at very close proximity to the CMB Cold Spot. While that can be coincidental, it can also indicate on a certain link between the CMB distribution and the distribution of galaxy spin directions. The nature of the CMB Cold Spot is still a mystery. It is statistically significant \citep{cruz655non}, consistent across different instruments, and cannot be explained by foreground contamination \citep{vielva2010comprehensive}, but there is still no clear explanation to its existence. One possible explanation is a supervoid in that part of the Universe \citep{masina2009cold}, but observations have shown no evidence of unusual distribution of galaxy population around the location of the CMB Cold Spot \citep{granett2010galaxy,mackenzie2017evidence}. Here, the CMB Cold Spot is aligned with an axis formed by the distribution of the spin directions of spiral galaxies. It should be mentioned that a link between cosmic vacuum and galaxy rotation has also been proposed \citep{chechin2010cosmic}.

The concept of a cosmological-scale axis has been proposed through theories related to the geometry of the Universe such as ellipsoidal universe \citep{campanelli2006ellipsoidal,campanelli2007cosmic,gruppuso2007complete,campanelli2011cosmic,cea2014ellipsoidal}. An ellipsoidal universe is not expected to be isotropic, and the anisotropty is expected to exhibit itself in the form of cosmological-scale quadrupole \citep{rodrigues2008anisotropic}. Another cosmological model the relies on the existence of a cosmological-scale axis is the rotating universe \citep{gamow1946rotating,godel1949example,ozsvath1962finite,ozsvath2001approaches,su2009universe,sivaram2012primordial,chechin2016rotation,chechin2017does,camp2021}. The existence of a cosmological-scale axis has also been linked to theories such as holographic big bang \citep{pourhasan2014out,altamirano2017cosmological}.

Black hole cosmology \citep{pathria1972universe,easson2001universe,seshavatharam2010physics,poplawski2010radial,chakrabarty2020toy} can also explain the existence of a cosmological-scale axis. Since stars spin, black holes also spin based on the spin of the stars from which they were created \citep{mcclintock2006spin}. If the Universe is hosted in a black hole, the Universe should have a preferred direction inherited from its host black hole \citep{poplawski2010cosmology,seshavatharam2010physics,seshavatharam2020light}, which would exhibit itself in the form of an axis \citep{seshavatharam2020integrated}. Such black hole universe might not be aligned with the cosmological principle \citep{stuckey1994observable}, but can explain other observations such as dark energy and the agreement between the Hubble radius and the cosmological Schwarzschild radius. 

A possible universal pattern of galaxy spin directions can be related to the proposed existence of a Universal force field \citep{barghout2019analysis}. The observation that galaxies in opposite lines of sight show opposite spin directions also agrees with cosmology driven by longitudinal gravitational waves \citep{mol2011gravitodynamics}, according which each galaxy at a certain distance from Earth is expected to have an antipode galaxy under the same physical conditions, but accelerating oppositely \citep{mol2011gravitodynamics}. This model also agrees with the greater asymmetry observed in the earlier Universe \citep{shamir2020patterns}.

The ability to analyze a possible non-random distribution of the spin directions of spiral galaxies is a research question that its studying was not practical in the pre-information era. As evidence for such non-random distribution are accumulating, additional research will be needed to fully understand its nature, and match it with other probes in addition to CMB.



\section*{Data Availability Statement}

Dataset from SDSS and HST used in this study are available freely from the URLs specified in the manuscript. Other datasets generated during this study are available on reasonable request.

\section*{Acknowledgments}

I would like to thank the knowledgeable referee for the useful comments that helped to improve the paper. The research was supported in part by NSF grants AST-1903823 and IIS-1546079.

The research is based on observations made with the NASA/ESA Hubble Space Telescope, and obtained from the Hubble Legacy Archive, which is a collaboration between the Space Telescope Science Institute (STScI/NASA), the Space Telescope European Coordinating Facility (ST-ECF/ESA) and the Canadian Astronomy Data Centre (CADC/NRC/CSA).

The Legacy Surveys consist of three individual and complementary projects: the Dark Energy Camera Legacy Survey (DECaLS; Proposal ID \#2014B-0404; PIs: David Schlegel and Arjun Dey), the Beijing-Arizona Sky Survey (BASS; NOAO Prop. ID \#2015A-0801; PIs: Zhou Xu and Xiaohui Fan), and the Mayall z-band Legacy Survey (MzLS; Prop. ID \#2016A-0453; PI: Arjun Dey). DECaLS, BASS and MzLS together include data obtained, respectively, at the Blanco telescope, Cerro Tololo Inter-American Observatory, NSF’s NOIRLab; the Bok telescope, Steward Observatory, University of Arizona; and the Mayall telescope, Kitt Peak National Observatory, NOIRLab. The Legacy Surveys project is honored to be permitted to conduct astronomical research on Iolkam Du’ag (Kitt Peak), a mountain with particular significance to the Tohono O’odham Nation.

NOIRLab is operated by the Association of Universities for Research in Astronomy (AURA) under a cooperative agreement with the National Science Foundation.

This project used data obtained with the Dark Energy Camera (DECam), which was constructed by the Dark Energy Survey (DES) collaboration. Funding for the DES Projects has been provided by the U.S. Department of Energy, the U.S. National Science Foundation, the Ministry of Science and Education of Spain, the Science and Technology Facilities Council of the United Kingdom, the Higher Education Funding Council for England, the National Center for Supercomputing Applications at the University of Illinois at Urbana-Champaign, the Kavli Institute of Cosmological Physics at the University of Chicago, Center for Cosmology and Astro-Particle Physics at the Ohio State University, the Mitchell Institute for Fundamental Physics and Astronomy at Texas A\&M University, Financiadora de Estudos e Projetos, Fundacao Carlos Chagas Filho de Amparo, Financiadora de Estudos e Projetos, Fundacao Carlos Chagas Filho de Amparo a Pesquisa do Estado do Rio de Janeiro, Conselho Nacional de Desenvolvimento Cientifico e Tecnologico and the Ministerio da Ciencia, Tecnologia e Inovacao, the Deutsche Forschungsgemeinschaft and the Collaborating Institutions in the Dark Energy Survey. The Collaborating Institutions are Argonne National Laboratory, the University of California at Santa Cruz, the University of Cambridge, Centro de Investigaciones Energeticas, Medioambientales y Tecnologicas-Madrid, the University of Chicago, University College London, the DES-Brazil Consortium, the University of Edinburgh, the Eidgenossische Technische Hochschule (ETH) Zurich, Fermi National Accelerator Laboratory, the University of Illinois at Urbana-Champaign, the Institut de Ciencies de l’Espai (IEEC/CSIC), the Institut de Fisica d’Altes Energies, Lawrence Berkeley National Laboratory, the Ludwig Maximilians Universitat Munchen and the associated Excellence Cluster Universe, the University of Michigan, NSF’s NOIRLab, the University of Nottingham, the Ohio State University, the University of Pennsylvania, the University of Portsmouth, SLAC National Accelerator Laboratory, Stanford University, the University of Sussex, and Texas A\&M University.

BASS is a key project of the Telescope Access Program (TAP), which has been funded by the National Astronomical Observatories of China, the Chinese Academy of Sciences (the Strategic Priority Research Program “The Emergence of Cosmological Structures” Grant \# XDB09000000), and the Special Fund for Astronomy from the Ministry of Finance. The BASS is also supported by the External Cooperation Program of Chinese Academy of Sciences (Grant \# 114A11KYSB20160057), and Chinese National Natural Science Foundation (Grant \# 11433005).

The Legacy Survey team makes use of data products from the Near-Earth Object Wide-field Infrared Survey Explorer (NEOWISE), which is a project of the Jet Propulsion Laboratory/California Institute of Technology. NEOWISE is funded by the National Aeronautics and Space Administration.

The Legacy Surveys imaging of the DESI footprint is supported by the Director, Office of Science, Office of High Energy Physics of the U.S. Department of Energy under Contract No. DE-AC02-05CH1123, by the National Energy Research Scientific Computing Center, a DOE Office of Science User Facility under the same contract; and by the U.S. National Science Foundation, Division of Astronomical Sciences under Contract No. AST-0950945 to NOAO.

SDSS-IV is managed by the Astrophysical Research Consortium for the Participating Institutions of the SDSS Collaboration including the Brazilian Participation Group, the Carnegie Institution for Science, Carnegie Mellon University, the Chilean Participation Group, the French Participation Group, Harvard-Smithsonian Center for Astrophysics, Instituto de Astrofisica de Canarias, The Johns Hopkins University, Kavli Institute for the Physics and Mathematics of the Universe (IPMU) University of Tokyo, the Korean Participation Group, Lawrence Berkeley National Laboratory, Leibniz Institut fur Astrophysik Potsdam (AIP), Max-Planck-Institut fur Astronomie (MPIA Heidelberg), Max-Planck-Institut fur Astrophysik (MPA Garching), Max-Planck-Institut fur Extraterrestrische Physik (MPE), National Astronomical Observatories of China, New Mexico State University, New York University, University of Notre Dame, Observatario Nacional / MCTI, The Ohio State University, Pennsylvania State University, Shanghai Astronomical Observatory, United Kingdom Participation Group, Universidad Nacional Autonoma de Mexico, University of Arizona, University of Colorado Boulder, University of Oxford, University of Portsmouth, University of Utah, University of Virginia, University of Washington, University of Wisconsin, Vanderbilt University, and Yale University.

The Pan-STARRS1 Surveys (PS1) and the PS1 public science archive have been made possible through contributions by the Institute for Astronomy, the University of Hawaii, the Pan-STARRS Project Office, the Max-Planck Society and its participating institutes, the Max Planck Institute for Astronomy, Heidelberg and the Max Planck Institute for Extraterrestrial Physics, Garching, The Johns Hopkins University, Durham University, the University of Edinburgh, the Queen's University Belfast, the Harvard-Smithsonian Center for Astrophysics, the Las Cumbres Observatory Global Telescope Network Incorporated, the National Central University of Taiwan, the Space Telescope Science Institute, the National Aeronautics and Space Administration under Grant No. NNX08AR22G issued through the Planetary Science Division of the NASA Science Mission Directorate, the National Science Foundation Grant No. AST-1238877, the University of Maryland, Eotvos Lorand University (ELTE), the Los Alamos National Laboratory, and the Gordon and Betty Moore Foundation.

\bibliographystyle{apalike}

\bibliography{asymmetry_combined}

\end{document}